\documentclass[12pt, a4]{article}

\usepackage{graphicx, hyperref}
\usepackage{amsfonts, amsmath,amssymb, mathrsfs, wasysym}
\usepackage[english]{babel}
\usepackage{datetime}
\usepackage[normalem]{ulem} 
\usepackage{color,slashed}

\ifx\pdfoutput\undefined
\usepackage[dvips,bookmarks]{hyperref}	
\else
\usepackage{hyperref}	
\fi
\hypersetup{colorlinks,bookmarksopen,bookmarksnumbered,citecolor=rossos,
linkcolor=rossos,pdfstartview=FitH,urlcolor=rossos}
\newcommand{\Tr}{{\rm Tr}\,}

\def\circa#1{\,\raise.3ex\hbox{$#1$\kern-.75em\lower1ex\hbox{$\sim$}}\,}

\numberwithin{equation}{section} 
\usepackage[table]{xcolor}

\definecolor{grigino}{cmyk}{0,0,0,0.2}
\definecolor{mentuccia}{cmyk}{0.4,0,0.3,0.1}
\definecolor{arancino}{cmyk}{0,0.1,0.4,0}
\definecolor{menta}{cmyk}{0.7,0,0.5,0.3}
\definecolor{grigios}{cmyk}{0,0,0,0.5}
\definecolor{bianco}{cmyk}{0,0,0,0}
\definecolor{arancio}{cmyk}{0,0.2,0.6,0}
\definecolor{grigio}{cmyk}{0,0,0,0.1}
\definecolor{rosa}{cmyk}{0,0.1,0.1,0.02}
\definecolor{rosino}{cmyk}{0,0.05,0.05,0.02}
\definecolor{rosas}{cmyk}{0,0.3,0.25,0.05}
\definecolor{celeste}{cmyk}{0.1,0,0,0.02}
\definecolor{giallino}{cmyk}{0,0,0.4,0.02}
\definecolor{rosso}{cmyk}{0,1,1,0.4}
\definecolor{rossos}{cmyk}{0,1,1,0.55}
\definecolor{rossoc}{cmyk}{0,1,1,0.2}
\definecolor{blu}{cmyk}{1,1,0,0.3}
\definecolor{blus}{cmyk}{1,1,0,0.5}
\definecolor{bluc}{cmyk}{1,1,0,0.1}
\definecolor{blucc}{cmyk}{0.7,0.5,0,0}
\definecolor{viola0}{cmyk}{0,0.4,0,0.04}
\definecolor{viola}{cmyk}{0,0.5,0,0.05}
\definecolor{viola2}{cmyk}{0,1,0.2,0.6}
\definecolor{verde}{cmyk}{0.92,0,0.59,0.25}
\definecolor{verdec}{cmyk}{0.92,0,0.59,0.15}
\definecolor{verdecc}{cmyk}{0.42,0,0.8,0.05}
\definecolor{verdes}{cmyk}{0.92,0,0.59,0.4}
\definecolor{verdino}{cmyk}{0.12,0,0.3,0.02}
\definecolor{giallo}{cmyk}{0,0,1,0}
\definecolor{gialloverde}{cmyk}{0.44,0,0.74,0}

\newcommand{\tab}{\,,\quad}
\newcommand{\nn}{\nonumber}
\newcommand{\be}{\begin{equation}}
\newcommand{\ee}{\end{equation}}
\newcommand{\bea}{\begin{eqnarray}}
\newcommand{\eea}{\end{eqnarray}}
\newcommand{\gev}{\textrm{ GeV}}
\newcommand{\tev}{\textrm{ TeV}}

\newcommand{\lgm}{lopsided gauge mediation }

\newcommand{\newc}{\newcommand}

\newc{\gsim}{\lower.7ex\hbox{$\;\stackrel{\textstyle>}{\sim}\;$}}
\newc{\lsim}{\lower.7ex\hbox{$\;\stackrel{\textstyle<}{\sim}\;$}}

\numberwithin{equation}{section}

\begin{document}

\date{\mbox{}}

\title{
\vspace{-2.0cm} 
\begin{flushright}
\normalsize{CERN-PH-TH/2011-066 }\\
\end{flushright}
\vspace{2.0cm}
{\bf \Huge Lopsided Gauge Mediation} \\[8mm]}

\author{Andrea De Simone $^a$, Roberto Franceschini $^a$, Gian Francesco Giudice $^b$,\\
Duccio Pappadopulo $^a$, Riccardo Rattazzi $^a$.\\[8mm]
\normalsize\it $^a$ Institut de Th\'eorie des Ph\'enom\`enes Physiques, EPFL,  CH--1015 Lausanne, Switzerland\\[-0.05cm] 
\normalsize\it $^b$ CERN, Theory Division, CH-1211 Geneva 23, Switzerland\\[-0.05cm]}

\maketitle

\setcounter{page}{0}
\thispagestyle{empty}

\begin{abstract}
\noindent
It has been recently pointed out that the unavoidable tuning among supersymmetric parameters required to raise the Higgs boson mass beyond its experimental limit opens up new avenues for dealing with the so called $\mu$-$B_\mu$ problem of gauge mediation. In fact, it allows for accommodating, with no further parameter tuning, large values of $B_\mu$ and of the other Higgs-sector soft masses, as predicted in models where both $\mu$ and $B_\mu$ are generated at one-loop order. This class of models, called Lopsided Gauge Mediation, offers an interesting alternative to conventional gauge mediation and is characterized by a strikingly different phenomenology, with light higgsinos, very large Higgs pseudoscalar mass, and moderately light sleptons. 
We discuss general parametric relations involving the fine-tuning of the model and various observables such as the chargino mass and the value of $\tan\beta$. We build an explicit model and we study the constraints coming from LEP and Tevatron. We show that in spite of new interactions between the Higgs and the messenger superfields, the theory can remain perturbative up to very large scales, thus retaining gauge coupling unification. 
\end{abstract}

\newpage

\def\eq#1{eq.~(\ref{#1})}

\section{Introduction}
\label{Introduction}

One of the most appealing features of gauge mediation~\cite{dinenelson,gaugemediation} is the calculability of the soft terms. However, this predictive power is deficient unless one defines the mechanism that generates the $\mu$ and $B_\mu$ terms. This mechanism is problematic in gauge mediation because, on fairly general grounds, one obtains the relation~\cite{Dvali:1996cu}
\be
B_\mu = \mu \Lambda ,
\label{mubmu}
\ee
where $\Lambda$ is the effective scale of supersymmetry breaking. Since gauge mediation predicts $\Lambda \sim (16 \pi^2/g^2) m_{\rm soft}$, where $m_{\rm soft}$ is the characteristic size of soft masses and $g$ is a gauge coupling constant,
\eq{mubmu} leads to a one-loop mismatch between $B_\mu$ and $\mu^2$. Many authors have addressed this problem of gauge mediation by proposing solutions that circumvent \eq{mubmu}; see {\it e.g.} refs.~\cite{dinenelson,Dvali:1996cu,musol,csakietal,Komargodski:2008ax}.

The problematic relation in \eq{mubmu} follows from the fact that both $\mu$ and $B_\mu$ originate
at the same order in perturbation theory upon integrating out the messengers. For instance, in the simplest case where the messenger threshold is controlled by a spurion superfield $X=M(1+\Lambda\theta^2)$, by dimensional analysis the generic finite 1-loop correction to the K\"ahler potential has the form
\be
c \int d^{4}\theta\, H_{u}H_{d}F(X/X^{\dagger}),
\label{opop}
\ee
giving rise to  $\mu \propto B_\mu\propto F'$. 
Here $c$ is typically a one-loop factor, but \eq{mubmu} stems from \eq{opop}, independently of the value of $c$. This indeed shows that the $\mu$--$B_\mu$ problem, defined by \eq{mubmu}, holds in a more general context than just gauge mediation and it potentially exists in any theory in which there is a separation of scales between $\Lambda$ and $m_{\rm soft}$. Such a separation is often needed in order to address the flavor problem and it is present in theories with anomaly mediation~\cite{anowww} or gaugino mediation~\cite{gaugwww}\footnote{In those models $F(X/X^\dagger)$ in eq.~(\ref{opop}) is replaced by a generic function of $X$ and $X^\dagger$, but the conclusion are similar. In the simplest gaugino mediated models the small parameter $c$ is proportional to the inverse of the volume of the compact extra dimension through which supersymmetry breaking is mediated.}. The relation \eq{mubmu} becomes satisfactory only when $\Lambda$ is of the order of the soft masses (as in gravity
mediation~\cite{Giudice:1988yz}). Thus, on rather general grounds, it seems that supersymmetric model building is facing a clash between natural solutions to the flavor problem and to the $\mu$--$B_\mu$ problem.

 Since the experimental searches for charginos constrain $\mu >100$~GeV, the relation $B_\mu \sim (4\pi \mu)^2/g^2$ implies an anomalously large value of $B_\mu$. This large value of $B_\mu$ either destabilizes the Higgs potential along the D-flat direction or requires a considerable parameter tuning in order to adjust the Higgs vacuum expectation value. As eliminating any source of tuning in the Higgs sector is the very reason why low-energy supersymmetry was originally introduced, the $\mu$--$B_\mu$ problem of gauge mediation cannot be ignored.

The situation about fine tuning in supersymmetry has changed significantly in the last decade. The unsuccessful searches for supersymmetric particles and, especially, for a light Higgs boson have constrained the models to such a degree that tunings at the level of a few percent seem almost inescapable. The problem is particularly acute in models such as gauge mediation, where the soft terms are tightly determined by the theory and cannot satisfy some special, but propitious, relations capable of partly alleviating the tuning problem. If gauge mediation is realized in nature, it must endure some accidental tuning.

The latter consideration prompts us to revisit the viability of \eq{mubmu}. Our study is directly motivated by the analysis in ref.~\cite{csakietal}, on which we shall elaborate.  The main observation is that, once one implements the tuning necessary to evade the Higgs-mass limit from LEP, \eq{mubmu} does not necessarily imply any further tunings in the theory. In certain cases that will be studied in this paper, a single tuning is sufficient to take care of both the Higgs mass and the anomalously large $B_\mu$. This class of models will be called here {\it lopsided gauge mediation}.

An important difference of our study with respect to the analysis of ref.~\cite{csakietal} is that
 we shall insist on calculability (weak coupling)  and  perturbative unification. In particular we shall work out an explicit model overcoming a quantitative problem of the minimal example of ref. \cite{Dvali:1996cu}.
  It is interesting that the assumption of calculability renders lopsided gauge mediation  viable only in the presence of a separation between $m_Z$ and the sparticle masses. In particular it becomes viable in the presence of the well known tuning that is necessary to lift the Higgs mass above the LEP bound. In this respect, we must stress  that 
 lopsided gauge mediation is not ``natural": a fine tuning is actually present. However, this tuning simply amounts to what is required in order to lift the Higgs mass beyond its present experimental limit. It corresponds to the  inevitable tuning present in any model of gauge mediation.

In spite of the necessary tuning, one cannot regard lopsided gauge mediation inferior to the ordinary gauge mediation scheme, in which $\mu$ and $B_\mu$ of the order of the weak scale are put in by hand or generated by new interactions at some intermediate scale. Lopsided gauge mediation is equally good (or equally bad) since the degree of fine tuning is identical. There is however an important difference with respect to the ordinary scheme of gauge mediation, since the phenomenology is totally distinct. Lopsided gauge mediation is characterized by a very small $\mu$, leading to light higgsinos, by a pseudoscalar-Higgs mass $m_A$ in the range of several TeV and a moderate to large value of $\tan\beta$. The phenomenology of lopsided gauge mediation, discussed in this paper, is worth serious scrutiny by the LHC experimental collaborations, since it represents just another side of gauge mediation, equally likely from the theoretical point of view, but much less explored.

\section{Characterizing Lopsided Gauge Mediation \label{pattern}}

The condition for electroweak symmetry breaking in supersymmetry, in the presence of an unavoidable 
hierarchy between
$m_Z$ and $m_{\rm soft}$, can be expressed as a requirement of near criticality: in the phase diagram spanned by the model parameters, the soft terms have to lie very close to the line separating the broken and unbroken phases of $SU(2)\times U(1)$~\cite{Giudice:2006sn}. Indicating by ${\cal M}^2_H$  the Higgs mass matrix at vanishing background field value, the condition of near criticality is just ${\rm det}\, {\cal M}^2_H \simeq 0$. For sufficiently low messenger scale one has 
\be
{\cal M}_H^2=
\left(
\begin{matrix}
m_{H_u}^2-\delta m_{H_u}^2+|\mu|^2&B_\mu\\
B_\mu& m_{H_d}^2+|\mu|^2
\end{matrix}
\right)
\label{massmatrix}
\ee
where $m_{H_{u,d}}^{2}$ are the usual Higgs soft masses evaluated at the messenger scale and $\delta m_{H_u}^2$ is the radiative correction proportional to the stop square mass
\be
\delta m_{H_u}^2\simeq {3\alpha_t \over \pi} m_{\tilde t}^2 \log {M\over m_{\tilde t}}.\label{dmhuq}
\ee
The condition of criticality then becomes
\be
\left( m_{H_{u}}^{2} -\delta m_{H_u}^2 +|\mu|^2\right)
\left( m_{H_{d}}^{2}  +|\mu|^2\right)\simeq B_\mu^2\, .
\label{crit}
\ee
The stop induced correction $\delta m_{H_u}^2$ is large compared to $M_Z^2$ because the stop mass has to be close to 1~TeV to lift the Higgs boson mass above the LEP limit. The largeness of this term is at the origin of the usual fine tuning problem of supersymmetry. As evident from \eq{crit}, there are two extreme ways of implementing the tuning required by near criticality: the negative radiative correction can be compensated either by a large $\mu^2$ or by a large $m_{H_{u}}^2$. The first option corresponds to ordinary gauge mediation; the second one to lopsided gauge mediation. Let us consider the two cases.

{\it (i)} In ordinary gauge mediation, one takes $\mu^2 \simeq \delta m_{H_u}^2$ and $m_{H_{u,d}}^2\ll \mu^2$. Electroweak breaking is achieved at the price of a tuning involving a large value of $\mu$. In this case, from the condition of criticality we find
\be
\label{casei}
\frac{B_\mu}{\mu^2} \simeq \left( 1-\frac{\delta m_{H_u}^2}{\mu^2}\right)^{1/2} <1.
\ee
This result is clearly incompatible with \eq{mubmu} which requires $B_\mu /\mu^2 \sim 16 \pi^2$. Ordinary gauge mediation suffers from a $\mu$--$B_\mu$ problems and requires a special solution.

{\it (ii)} In lopsided gauge mediation, we choose $m_{H_u}^2 \simeq \delta m_{H_u}^2$ and $\mu^2 \ll m_{H_{u,d}}^2$. The condition of criticality now becomes
\be
\frac{B_\mu}{\mu^2} \simeq \left( 1-\frac{\delta m_{H_u}^2}{m_{H_u}^2}\right)^{1/2} \frac{m_{H_u}m_{H_d}}{\mu^2}.
\ee
This result is not incompatible with \eq{mubmu}, because $B_\mu/\mu^2$ can be as large as $16\pi^2$, as long as $m_{H_d}^2$ is sufficiently large. Note that the largeness of $m_{H_d}^2$ in the mass matrix
(\ref{massmatrix}) 
does not necessarily imply a further tuning for the theory, if $m_{H_u}^2m_{H_d}^2\propto B_\mu^2$. We will show in the next section that the simplest model of lopsided gauge mediation automatically predicts this relation of proportionality. The issue about fine tuning will be addressed in sect.~\ref{tuning}.

\section{A Higgs sector coupled to the messengers 
\label{setup}}
%
In order to set the stage and fix the notation let us recall the basics of minimal gauge mediation.
The masses and splittings of the messengers ($\overline\Phi$, $\Phi$) are all controlled by a spurion superfield  $X=M(1+\Lambda_G\theta^2)$
via  the superpotential
\be
W=X\overline\Phi\Phi\,.
\ee
The dynamics that gives rise to $X\not = 0$ arises in an unspecified ``secluded sector", which contains also the supersymmetry breaking dynamics. On general grounds one should interpret $X$ as the expectation value 
\be
X=M+k\langle X_{NL}\rangle
\ee
where  $X_{NL}=F(\theta +\chi/{\sqrt 2F})^2$ is the canonical Goldstino superfield~\cite{seiberg}, while  $k$  effectively describes the coupling of the messengers to the supersymmetry breaking dynamics. By naive dimensional analysis (NDA) one expects $k\lsim 4\pi$, though in several realistic models one has $k\ll 1$.
At the scale $M$ the soft masses are given by
\be
M_a= n_G \Lambda_G {\alpha_a\over 4\pi}\,,\quad \label{massesGM}
\tilde m^2_I=2 n_G \Lambda_G^2 \sum_{a=1}^3 C_a^{(I)} \left({\alpha_a\over 4\pi}\right)^2,
\ee
where $C^{(I)}$ are
the quadratic Casimirs of the representation $I$ of the SM gauge group. Here $n_G$ is defined as an effective messenger number which is, for instance, equal to $1$ if the messenger are a $\boldsymbol{5} \oplus\boldsymbol{\overline{5}}$ of $SU(5)$ or to 3 in the case they belong to a $\boldsymbol{10} \oplus\boldsymbol{\overline{10}}$. The minimal scenario is easily generalized  by promoting $M$ and $k$ to matrices under the unique constraint of $SU(3)\times SU(2)\times U(1)$ gauge invariance. Indeed in that way one covers a chunk of the full 6-dimensional parameter space of general gauge mediation \cite{ggm,Carpenter:2008wi}.

At this level $\mu$ and $B_\mu$ are vanishing while the masses $m_{H_u}$ and $m_{H_d}$ are equal at the scale $M$. A possible way to generate $\mu$ and $B_\mu$ is adding direct couplings of the Higgs doublets to two pairs of messenger fields 
\be
\label{extendedGMSB}
W=
\lambda_u H_u D S+\lambda_d H_d \bar D \bar S+
X_D D\bar D+X_S S \bar S.
\ee
The simplest choice for the messenger fields is taking $D$ and $\bar D$ to be $SU(2)$ doublets with hypercharge $\pm 1$ (being part of a complete GUT representation, such as the fundamental of $SU(5)$), while $S$ and $\bar S$ are weak and hypercharge (as well as GUT) singlets. In this case it is more economical to identify $S$ and $\bar S$ with a single chiral superfield. 
In \eq{extendedGMSB} $X_{D}$ and $X_S$ are spurions with both scalar and $F$ components which are conveniently parametrized as
\be\label{Xparam}
X_{D,S}=M_{D,S}(1+\Lambda_{D, S}\,\theta^2)\,.
\ee
Notice that all phases in eqs.~(\ref{extendedGMSB}) and (\ref{Xparam}) can be eliminated by field redefinitions up to the relative phases between the various supersymmetry-breaking masses $\Lambda_{G,D,S}$ (these parameters have indeed the same quantum numbers under all spurionic global symmetries). The remaining phases
explicitly break CP and dangerously contributes to edms. We shall thus assume that the secluded sector respects CP, so that $\Lambda_D$ and $\Lambda_S$ can also be chosen real.
After integrating out the messengers at one-loop, the low-energy K\"ahler potential is renormalized in a calculable fashion \cite{effectivekahler} and the soft parameters of the Higgs sector are easily calculated (see Appendix~\ref{app:kahler} for details)
\bea
m_{H_{u,d}}^2&=&{\lambda_{u,d}^2\over 16\pi^2}\Lambda_D^2\,  P (x,y)
\label{mHfromKahler}\\
\mu&=&{\lambda_u\lambda_d\over 16\pi^2}\Lambda_D\,  Q(x,y)
\label{muq}\\
B_\mu&=&{\lambda_u\lambda_d\over 16\pi^2}\Lambda_D^2 \, R(x,y)
\label{Bmur}\\
A_{u,d}&=&{\lambda^2_{u,d}\over 16\pi^2}\Lambda_D \, S(x,y)\, . 
\label{At}
\eea 
Here we have set $x=M_S/M_D, y=\Lambda_S/\Lambda_D$ and
defined the functions
\bea
\label{loopfunctions}
P(x,y)&=&{x^2(1-y)^2\over (x^2-1)^3}\left[2(1-x^2)+(1+x^2) \log x^2\right]\\
Q(x,y)&=&{x \over (x^2-1)^2} \left[(x^2-1)(1-y)+(y-x^2)\log x^2\right]\\
R(x,y)&=&{x  \over (x^2-1)^3}
\left\{ (1-x^4)(1-y)^2+\left[2x^2(1+y^2)-y(1+x^2)^2\right]\log x^2\right\} \\
S(x,y)&=&{1  \over (x^2-1)^2}
\left[(x^2-1)(1-x^2y)-x^2(1-y)\log x^2\right]\,.
\label{loopfunctions2}
\eea
The function $P$ is positive and thus the square masses $m_{H_{u,d}}^2$ are also positive. However, $P$ vanishes for $y=1$ ($\Lambda_D =\Lambda_S$). This is an instance of the known result that one-loop soft masses vanish in the presence of just one spurion $X_D\propto X_S$~\cite{Giudice:1997ni}. Notice also that, by the argument on the elimination of unphysical phases, it follows that the sign of $y$ is physical while the sign of $x$ is not. Consistently with that, one has that the rephasing invariant combination $A\mu B_\mu^*\propto QRS$ is a function of $x^2$ and $y$.
 Also worth noticing is that all the four functions in eqs.~(\ref{loopfunctions})--(\ref{loopfunctions2}) are smaller than one in absolute value, with $P$ being somewhat smaller than the others. An important feature is that $R$ and $P$ satisfy $|R(x,y)|\geq 2P(x,y)$ for any $x$ and $y$.
This relation is problematic because at the messenger scale (where $\delta m_{H_u}^2=0)$ we find that, for small $\mu$,
\be
{\rm det}\, {\cal M}^2_H \simeq m_{H_u}^2m_{H_d}^2-B_\mu^2\propto P^2(x,y)-R^2(x,y).
\ee
Since $R^2$ is always bigger than $P^2$, one has ${\rm det}\, {\cal M}^2_H<0$ so  that  electroweak symmetry is  broken at the large scale $M$, leading to a spectrum with far too light sparticles.

It is clear that it must be possible to overcome the above difficulty by a slight complication of the model. This is because $B_\mu\propto R$ possesses $U(1)_R$ charge, so that one could conceive of contributions with opposite signs due to the presence of several spurions. Indeed it is enough to consider two gauge singlets $S$ and $\bar S$, but with a superpotential featuring  additional mass terms with respect to  eq.~(\ref{extendedGMSB})
\be\label{newmodel}
W=\lambda_u H_u D S+\lambda_d H_d \bar D \bar S+X_D D\bar D+ \frac{X_S}{2} \left( a_{S} S^2+a_{\bar S} {\bar S}^2+2a_{S\bar S}S\bar S\right)\,.
 \ee
The generation of a non-vanishing contribution to both $\mu$ and $B_\mu$ is related to the $a_{S\bar S}$ coefficient. 
If $a_{S\bar S}=0$ the $Z_2$ symmetry under which $H_u\rightarrow -H_u$ and $S\rightarrow -S$ forbids $\mu$ and $B_\mu$. After
diagonalizing the mass term in eq.~(\ref{newmodel}) and rescaling $X_S$, the superpotential becomes
\be\label{2singletsmodel}
W=\lambda_u H_u D(c_\theta S+s_\theta \bar S)+\lambda_d H_d \bar D (-s_\theta S+c_\theta \bar S)+X_D D\bar D+ \frac{X_S}{2} \left( S^2+\xi {\bar S}^2\right)
\,,
\ee
where $s_\theta\equiv \sin \theta$, $c_\theta\equiv \cos \theta$ and
\be\label{prrr}
\tan 2\theta =\frac{2a_{S\bar S}}{a_{\bar S}-a_S}\, ,~~~~~~\xi =\frac{a_{\bar S}-a_S\tan^2\theta}{a_S-a_{\bar S}\tan^2\theta}\,.
\ee
Again we have assumed CP invariance and chosen all parameters real.
The boundary conditions for the soft parameters in the Higgs sector are the sum of the $S$ and $\bar S$ contributions
\bea\label{boundary}
m_{H_{u}}^2&=&{|\lambda_{u}|^2\over 16\pi^2}\Lambda_D^2\, 
\left[c_\theta^2\, P(x,y)+s_\theta^2\, P(\xi x,y)\right]
\equiv a_1^{u}\dfrac{|\lambda_{u}|^2}{16\pi^2}\Lambda_D^2
\label{boundarymHu}\\
m_{H_{d}}^2&=&{|\lambda_{d}|^2\over 16\pi^2}\Lambda_D^2\, 
\left[ s_\theta^2\,P(x,y)+ c_\theta^2\,P(\xi x,y)\right]
\equiv a_1^{d}\dfrac{|\lambda_{d}|^2}{16\pi^2}\Lambda_D^2
\label{boundarymHd}\\
\mu&=&{\lambda_u\lambda_d\over 16\pi^2}\Lambda_D\, s_\theta c_\theta
\left[ - Q(x,y)+ Q(\xi x,y)\right]
\equiv a_2\dfrac{\lambda_u\lambda_d}{ 16\pi^2}\Lambda_D
\label{boundarymu}\\
B_\mu&=&{\lambda_u\lambda_d\over 16\pi^2}\Lambda_D^2 \,
s_\theta c_\theta \left[- R(x,y)+ R(\xi x,y)\right]
\equiv a_{3}\dfrac{\lambda_u\lambda_d}{16\pi^2}\Lambda_D^2
\label{boundaryBmu}\\
A_{u}&=&{|\lambda_{u}|^2\over 16\pi^2}\Lambda_D\, 
\left[ c_\theta^2\, S(x,y)+s_\theta^2\, S(\xi x,y)\right]\\
A_{d}&=&{|\lambda_{d}|^2\over 16\pi^2}\Lambda_D\, 
\left[s_\theta^2\,S(x,y)+ c_\theta^2\,S(\xi x,y)\right] \, .
\eea
From eq.~(\ref{boundaryBmu}) it is clear that we can now avoid the relation that lead to a negative ${\rm det}\, {\cal M}^2_H$. Indeed,
the $B_\mu$ term can be made arbitrarily small either by approaching the $Z_2$ symmetric limit ($\theta \to 0$), or by taking $\xi \to 1$, such that the electroweak critical condition in \eq{crit} can be satisfied. In these limits also $\mu$ will be suppressed.

\subsection{Naturally vanishing $B_\mu$}

The remark leading to the last model is that  $B_\mu$, unlike diagonal masses, transforms under $U(1)_R$ phase rotations. Then its overall relative size must  basically be a free parameter, given sufficient freedom to pick a model. Taking the same argument to its extreme we could indeed conceive a model in which $U(1)_{PQ}$ is broken in the sector that couples $H_1$ and $H_2$ to the messengers,
 while $U(1)_R$ is not. Of course in order to generate gaugino masses $U(1)_R$ must be broken by other messengers which for some reason do not couple to the Higgses. An example of a Higgs-messenger sector with the desired features is offered by a variation of the models above with the simple addition of an extra pair $D'$, $\overline D'$ of doublets, with the identification of $S$ and $\bar S$, and with superpotential
\be\label{bmu0}
W=
\lambda_u H_u D S+\lambda_d H_d \bar D S+
X D\bar D+M_S  S^2 + M_D (D\bar D'+D'\bar D).
\ee
with $X\equiv \langle X_{NL}\rangle =F\theta^2$. The above superpotential is the most general one compatible with the $R$ and $PQ$-charges in table \ref{charges}. 
\begin{table}[h]
\centering
\begin{tabular}{cccccccccccc}
\hline
&$H_u$& $H_d$& $D$ & $\overline D$ & $S$& $\overline S$&$D'$&$\overline{D'}$&X\\
\hline
$R$&1& 1& 0 & 0 & 1& 1&2&2&2\\
$PQ$&1& 1& -1 & -1 & 0& 0&1&1&2\\
\hline
\end{tabular}
\caption{\small $R$ and $PQ$ charge assignments for the superpotential in eq.~(\ref{bmu0}).}
\label{charges}
\end{table}
The expectation value of $X$ breaks PQ. On the other hand, since $X$ has R-charge 2, $F$ is $R$-neutral.  $R$-symmetry is thus exact in the above lagrangian. On the other hand the expectation value of $X$ breaks PQ, and one can easily verify (see Appendix \ref{app:kahler}) that at one-loop the renormalization of the Kahler potential 
gives rise to non-vanishing $\mu \sim F/M$,  $m_{H_u}^2$ and $m_{H_d}^2$,  while $B_\mu=0$.  The model contains an axion, which could be interpreted as the benign invisible axion if the scale of supersymmetry breaking is sufficiently high. Alternatively, the would-be axion could eliminated from the low-energy spectrum if PQ is only an approximate symmetry of the Higgs-messenger sector, as in the case of the $R$ symmetry.   

It remains an issue of model building to come up with a fundamental theory where $U(1)_R$ emerges as an accidental symmetry of the Higgs-messenger sector, while it is broken elsewhere. It does not seem implausible that it could happen, although we have not investigated specific implementations. Notice also that 
our mechanism, differs from the simple addition of $\mu H_1 H_2$ in the superpotential, which also breaks $U(1)_{PQ}$ and preserves $U(1)_R$, in that the origin of $\mu$ is here tied to supersymmetry breaking. This seems more satisfactory as it makes  $\mu\sim m_{soft}$ more plausible. On the other hand, in our case one necessarily has 1-loop contributions to $m_{H_u}^2$ and $m_{H_d}^2$. Indeed in order to preserve $U(1)_R$ the messenger masses in eq.~(\ref{bmu0}) must be controlled by multiple spurions $X, M_S, M_D$  with different global quantum numbers. Then one cannot rely on the usual theorem insuring the vanishing of 1-loop masses in the presence of a single spurion field controlling the whole spectrum and its mass splittings. 
The role of $U(1)_R$ in protecting $B_\mu$ was already emphasized in ref. \cite{Komargodski:2008ax}. However, following that remark, in the models of that paper $\mu$ 
originates from the superpotential rather than from the K\"ahler potential, and is thus  not directly related to supersymmetry breaking.

The necessary appearance of $m_{H_u}^2$ and $m_{H_d}^2$ at 1-loop makes the above model a special case of lopsided gauge-mediation: all the relations discussed in the previous section (such as $\mu^2\ll m_{H_u}^2< m_{H_d}^2$) hold true, but with the additional property that $B_\mu$ vanishes at the messenger scale.  Similarly to the model discussed in ref.~\cite{Rattazzi:1996fb},
$B_\mu$ is generated dominantly via RG running, leading to a naturally large value of $\tan\beta$ over a surprisingly large span of messenger masses.

\section{The maximal $\mu$}
\label{sect:mumax}

We can use eq.~(\ref{crit}) to get an approximate analytical understanding of how the electroweak breaking condition is satisfied. Dropping $\mu^2$, which is legitimate for perturbative values of the $\lambda$ couplings, we can solve for $\lambda_u$
\be\label{lambdau}
\frac{\lambda_u^2}{16\pi^2}\simeq\frac{1}{a_1^u(1-a_{B_\mu}^2)}\left[\frac{3\alpha_t m^2_{\tilde t}}{\pi \Lambda_D^2}
\log{ M \over m_{\tilde t}}-\frac{\Delta m_{H_u}^{2{\rm (GM)}}}{\Lambda_D^2}-\frac{\Delta m_{H_u}^{2{\rm (FI)}}}{\Lambda_D^2}\right],
\ee
where we defined $a_{B_\mu}^2=a_3^2/a_1^ua_1^d$.
%
Here $\Delta m_{H_u}^{2\rm{(GM)}}$, and $\Delta m_{H_u}^{2\rm{(FI)}}$  are weak-size contributions coming respectively from the standard gauge mediated part, see \eq{massesGM}, and the hypercharge Fayet-Iliopoulos contribution (see below) to $m_{H_u}^2$. Eq.~(\ref{lambdau}) turns out to be quite reliable, typically in the $\mathcal O(20\%)$ range,
if $a_{B_\mu}\neq 1$. For the semianalytic discussion of this session we shall however drop the second and third term within brackets in \eq{lambdau} since they are subdominant.
Note that under the reasonable assumption that $\Lambda_G \sim \Lambda_D$, one has $m_{\tilde t}^2/\Lambda_D^2\sim (g_3/4\pi)^4$. Thus,  as long as $a_{B_\mu}$ is not too close to 1, lopsided gauge mediation requires the one-loop expansion parameter $(\lambda_u/4\pi)^2$ to be roughly $(g_3/4\pi)^4$, that is two loops in QCD.

Once $\lambda_u$ is fixed, a lower bound on $\lambda_d$ is obtained from the experimental lower bound on the chargino mass, which can be approximately written as $\mu\gtrsim m_Z$. The limit can be expressed in terms of a dimensionless parameter $\eta=m_Z^2/m^2_{\tilde t}$ whose smallness quantifies the tuning of the supersymmetric model. Using (\ref{lambdau}) and (\ref{boundarymu}) we obtain
\be\label{eta1}
\frac{\lambda_d^2}{16\pi^2}\gtrsim \left[ \frac{a_1^u(1-a_{B_\mu}^2)\, \pi}{3a_2^2\, \alpha_t \, \log{ M \over m_{\tilde t}}}\right] \eta \, .
\ee
In our model the numerical pre-factor in front of $\eta$ can vary extensively throughout the parameter space. However, parametrically (and treating $\alpha_t \log{ M \over m_{\tilde t}}=O(1)$), \eq{eta1} corresponds to $\lambda_d \gtrsim 4\pi \eta^{1/2}$. Thus, the very existence of the tuning ({\it i.e.} the smallness of $\eta$) is the key element that allows for the existence of perturbative values of $\lambda_d$. The smaller is $\eta$, the wider is the range of $\lambda_d$ compatible with the experimental bound on $\mu$ and with a perturbative extrapolation of the theory up to high energies. 

The sizable $\lambda_d$ required by \eq{eta1} translates into a large value for $m_{H_d}^2$, which we recall arises already at 1-loop order unlike the sfermions masses. This is a trademark of lopsided gauge mediation and can have significant impact on the running of the soft masses through the induced hypercharge Fayet-Iliopoulos term (FI). This contribution is given by\footnote{In the formula we neglect the finite threshold corrections at the scale $m_{H_d}$. These will add to the logarithm a term $\mathcal O(1)$, which is negligible when $M$ is large enough.}
\be
\Delta m_{\tilde f}^{2{\rm (FI)}}\simeq-\frac{3Y_{\tilde f}\alpha_1 S}{10\pi}\, \log{ M\over m_{H_d}},
\ee
where
\be
S= 
{\rm Tr} (Y_{\tilde f}m^2_{\tilde f})= -m_{H_d}^2+m_{H_u}^2+  {\rm Tr} (m^2_Q-m^2_L-2m^2_U+m^2_D+m^2_E)
\,.
\ee
Such effects are large in lopsided gauge mediation and can easily reverse the ordering of sparticles masses predicted by ordinary gauge mediation.
For instance, we find that, for the first and second generation squark masses,$m_{u_{R}}^{2}<m_{d_{R}}^{2}$, a feature of the spectrum that is nevertheless hard to study at the LHC.

\begin{figure}[t]
  \begin{center}
    \includegraphics[width=8cm]{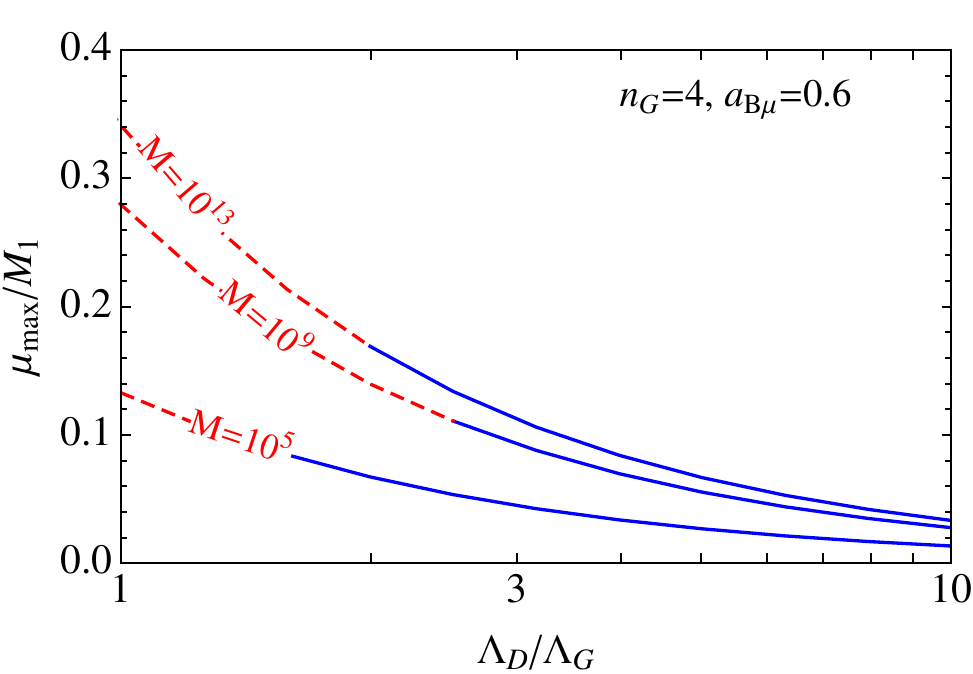}~~~~~~~
     \includegraphics[width=8cm]{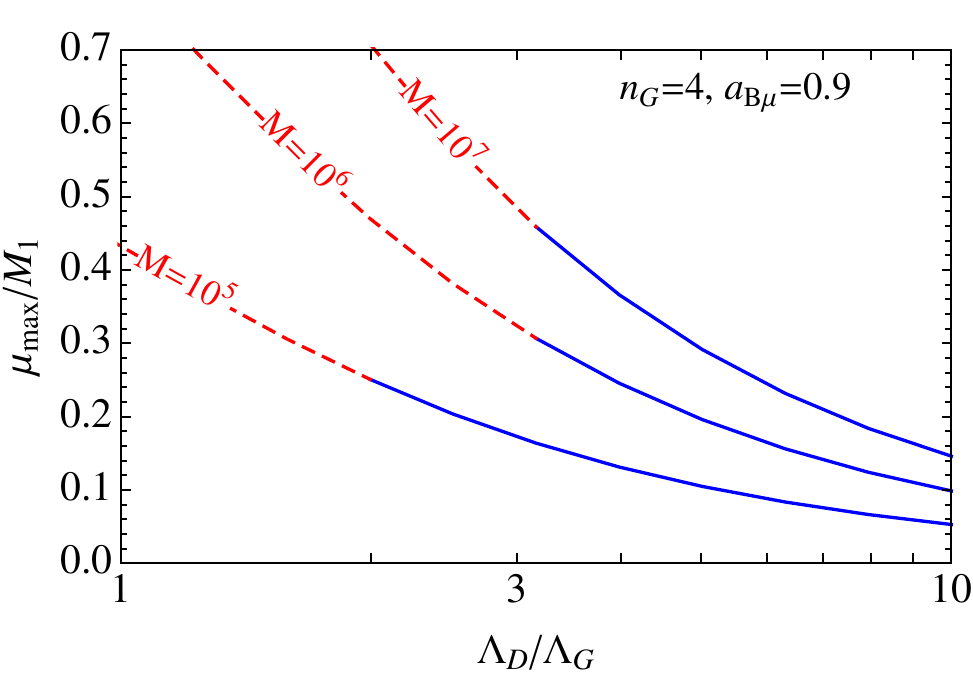}
  \end{center}
  \caption{\small $\mu_{\rm max}/M_1$ for a specific choices of parameters: $x=1.2,\,\xi x=4.5,\,\theta=0.8$ (left panel) and $x=0.7,\,\xi x=6.7,\,\theta=0.7$ (right panel). In both cases $y\ll1$ is assumed. On the red dashed branch of the curves the coupling $\lambda_d$ gets non perturbative at some point between $M$ and $M_{GUT}$.}
  \label{figmumax}
\end{figure}

A much more easily observable effect occurs among sleptons. Indeed the left-handed sleptons get a large and negative contribution to their square masses
\be
\Delta m_L^{2{\rm (FI)}}=-{3\alpha_1\over 20\pi}m_{H_d}^2 \log{ M\over m_{H_d}},\label{deltamslepton}
\ee
as opposed to the positive shift to the right-handed sleptons. 
The request of positive slepton square masses sets an upper bound on the value of $\lambda_d$ from eq.~(\ref{boundarymHd})
\be\label{lambdad}
\frac{\lambda_d^2}{16\pi^2}\lesssim\frac {5 \alpha_2^2\, n_G\Lambda_G^2}{8\pi \alpha_1 \, a_1^d\, \Lambda_D^2\, \log{ M \over m_{H_d}}}.
\ee
This can be combined with the value of $\lambda_u$ obtained from the criticality of electroweak breaking, eq.~(\ref{lambdau}), to get an upper bound for the value of $\mu$
\be\label{mumax}
\mu^2 \lesssim
\frac{ 15 a_\mu^2 \, \alpha_t \, \alpha_2^2\, n_G\Lambda_G^2\, m^2_{\tilde t}}
{8\pi^2(1-a_{B_\mu}^2)\alpha_1\, \Lambda_D^2} \, ,
\ee
where $a_\mu^2=a_2^2/a_1^ua_1^d$. From eq.~(\ref{mumax}) we observe that the ratio $\mu_{\rm max}/M_1$ is independent of $\Lambda_G$ (for fixed $\Lambda_G/\Lambda_D$), and only mildly dependent on both $n_G$ and $\log M$. The dependence on these latter parameters comes mainly from extra logarithms in $m_{\tilde t}^2$. For instance the positive gluino contribution to $m_{\tilde t}^2$ spoils the naive scaling with $n_G$ which is explicit in eq.~(\ref{mumax}): bigger values of $n_G$ results in bigger values of  $\mu_{\rm max}/M_1$. We show these effects in Fig.~\ref{figmumax}, for specific choices of the parameters. 
\newline

The value of $\mu$ can be further bounded from above requiring $\lambda_{d}$ to stay perturbative up to the GUT scale. 
The evolution of $\lambda_d$ is fully fixed only when the content of the messenger sector coupled to the Higgs fields is specified. In the the simplest case in which 
the messengers 
$D=(\boldsymbol{1},\boldsymbol{2})_{-1},~ S=  (\boldsymbol{1},\boldsymbol{1})_0$,
are  part of a fundamental and  a singlet of $SU(5)$ respectively,
 the RG equation for $\lambda_d$ is
\be
{d\lambda_d^2\over d t}={\lambda_d^2\over 8\pi^2}\left(
4\lambda_d^2-{3\over 5} g_1^2-3g_2^2
\right) \,,
\label{RGElambdad1}
\ee
with $t=\log Q$.
The significance of this perturbativity bound is shown in Fig.~\ref{figmumax}: on the red dashed part of the lines $\lambda_d$ hits the value $4\pi$ somewhere between $M$ and $M_{GUT}$. As the messenger scale is lowered, we expect a smaller value of $\mu$ because of the larger energy range between $M$ and $M_{GUT}$ in which the running coupling must remain perturbative\footnote{The perturbativity bound on $\lambda_d$ we discuss here requires qualifications. The
point is that, in our description of the models,  the dynamics that  breaks supersymmetry is not specified but simply parametrized effectively in terms of the spurions $X_D$, $X_S$. Somewhere below the GUT scale this effective description will have to be replaced by some physical interacting fields. This will in principle affect the RG evolution of our couplings, and of $\lambda_d$ in particular. However for the simple case in which the $X$ spurions originate from non-renormalizable operators suppressed by a scale larger than $ M_{GUT}$, there is clearly no effect on the RG evolution of $\lambda_{u,d}$. In this set up, assuming that the original scale of supersymmetry breaking is $\lsim 10^{10}$ GeV to suppress gravity mediated terms, one finds that hidden sector chiral operators of dimension  $d\leq 3$  could give rise to the $X$ spurions. The other possibility is that the $X$ spurions originate from $d=1$ chiral fields. In that case the presence of additional Yukawa interactions involving the messengers generally slows down the upward evolution of $\lambda_d$ and in principle relaxes our bound. On the other hand we do not expect that this effect can become dramatic without those other couplings also becoming large. We thus think that our upper bound on $\lambda_d$ has a broad validity.}.

\begin{figure}[t]
  \begin{center}
     \includegraphics[width=10cm]{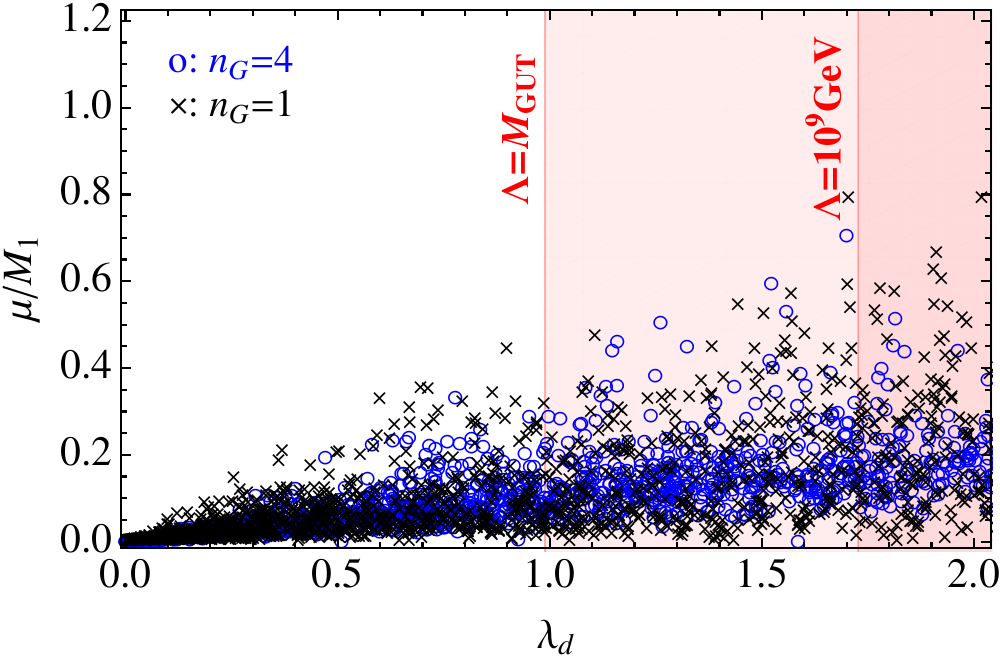}~~~~~~~
  \end{center}
  \caption{\small Scatter plot showing the values of $\mu/M_1$ attainable in the model described by eq.~(\ref{2singletsmodel}). The points are chosen randomly among those with $0.2\leq x\leq 4$, $4\leq \xi x \leq 10$, $0.5\leq\theta\leq 1$. This picks the most favorable region for the ratio $\mu/M_1$ as discussed in the text. Furthermore $y\ll 1$ is assumed, and $\Lambda_D/\Lambda_G$ varies between 0.1 and 10. The red shaded regions are those excluded by the perturbative requirement on $\lambda_d$ for a given choice of the cutoff scale of the theory: $M_{GUT}$ or $10^9$ GeV.}
  \label{scatmu}
\end{figure}

These general considerations imply an upper limit on $\mu/M_1$. The bound from perturbativity is generally stronger than the one imposed by the experimental limit on slepton masses, and this is particularly true if $\Lambda_D/\Lambda_G$ is not much bigger than one, see \eq{mumax}. The precise limit is ultimately related to the structure of the loop functions in eq.~(\ref{loopfunctions})--(\ref{loopfunctions2}). In the model of sect.~3 we find, quite generically, $\mu<M_1$. This is confirmed by the scatter plot for the ratio $\mu/M_1$ shown in the left panel of Fig.~\ref{scatmu} together with the region excluded by the perturbative requirement. All the points showed provide a viable spectrum, free of tachyons. We restricted the scanning of the parameter space to the most favorable region (see the caption in Fig.~\ref{scatmu}). Due to the cancellation in the boundary values of $\mu$ and $B_\mu$, see eqs.~(\ref{boundarymu})--(\ref{boundaryBmu}), bigger values for $\mu/M_1$ are obtained requiring a value of $\xi$ sufficiently far from 1. Furthermore $\theta\sim\pi/4$ is favored to avoid extra suppression in $\mu$.

The smallness of $\mu$, together with the mass relations imposed by gauge mediation, give significant constraints to the spectrum. Typically the higgsinos, the bino and the left-handed sleptons have masses close to $m_Z$, while the other superpartners are much heavier.
\newline

\begin{figure}[t]
  \begin{center}
     \includegraphics[width=8.5cm]{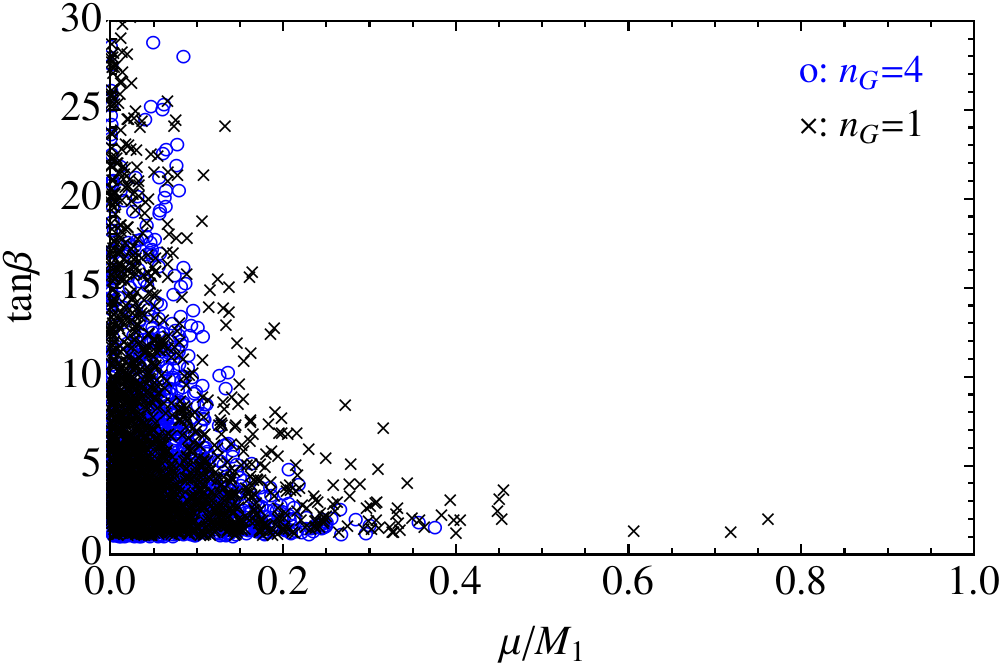}~~~~~~~
     \includegraphics[width=8.5cm]{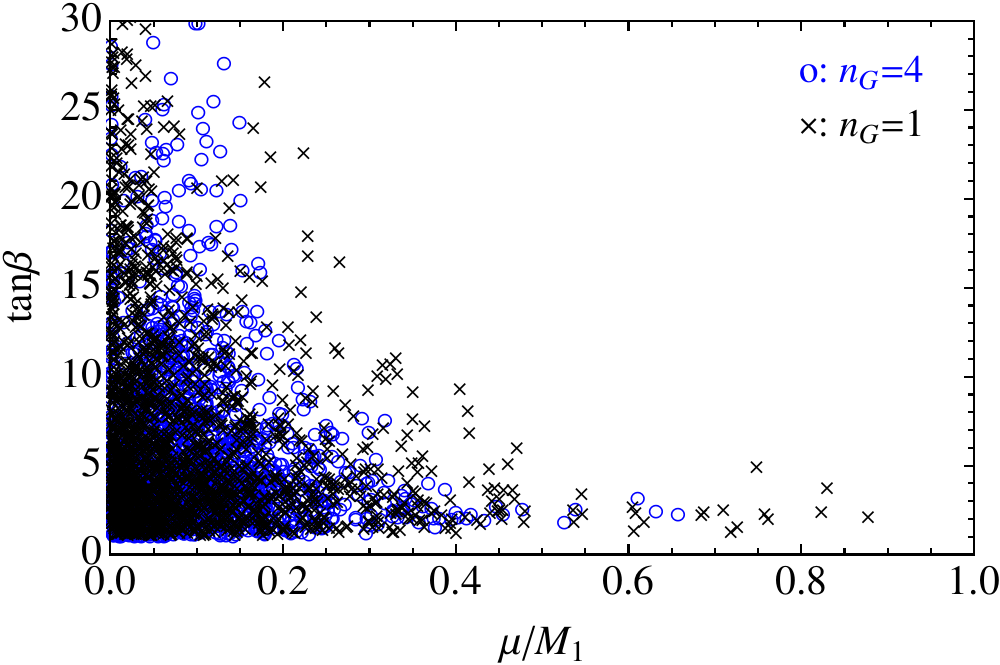}
  \end{center}
  \caption{\small Scatter plot showing the $\mu/M_1$-$\tan\beta$ correlation. The domain of the scatter-plot is the same as in Fig.~\ref{scatmu}. On the left we require perturbativity up to $\Lambda=M_{GUT}$ while on the right the cutoff is at $\Lambda=10^9$ GeV.}
  \label{scatmutan}
\end{figure}

\begin{figure}[t]
  \begin{center}
    \includegraphics[width=8cm]{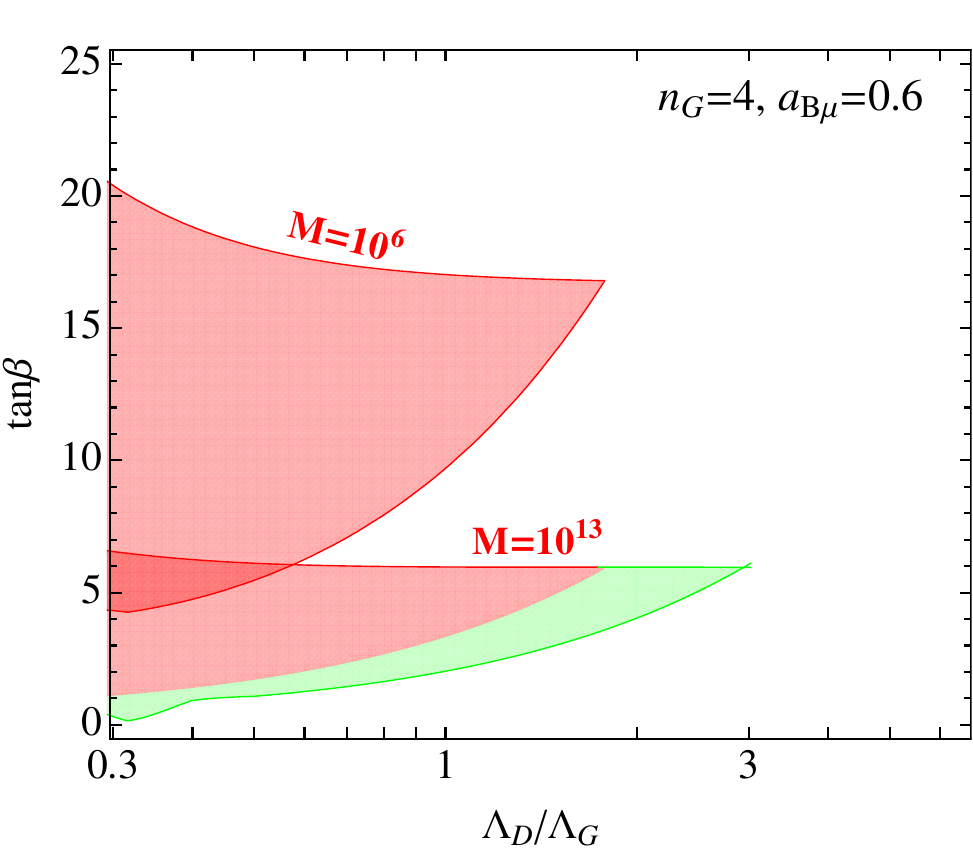}~~~~~~
    \includegraphics[width=8cm]{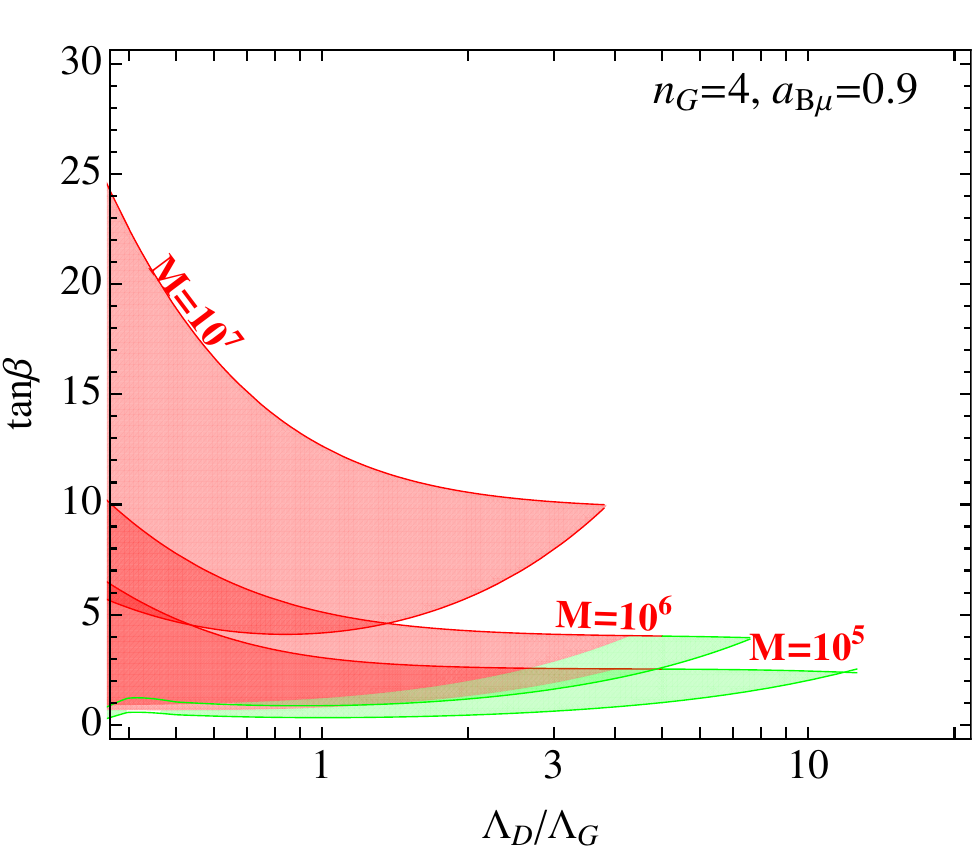}~~~~~~~~
  \end{center}
  \caption{
  \small Allowed values of $\tan\beta$ for a ratio $\eta=m_Z^2/m^2_{\tilde t}=1\permil$ as a function of $\Lambda_D/\Lambda_G$ for same choices of parameters as in Fig.~\ref{figmumax}. 
  On the red (dark) shaded regions the coupling $\lambda_d$ gets non perturbative at some point between $M$ and $M_{GUT}$.}
  \label{figtanmax}
\end{figure}

The value of $\tan\beta$ is is determined by eqs.~(\ref{boundarymHu})--(\ref{boundaryBmu}). Under the assumption of a hierarchy between $\lambda_u$ and $\lambda_d$, using $\sin 2 \beta=2B_\mu/( 2|\mu|^2+m_{H_u}^2+m_{H_d}^2)$, we obtain
\be
\label{tan2b} 
\tan\beta \simeq {a_1^d \lambda_{d} \over a_3   \lambda_{u} }.
\ee
The coupling $\lambda_u$ is fixed by the electroweak breaking condition as in eq.~(\ref{lambdau}), while the
coupling $\lambda_d$ is bounded both from above (slepton masses and perturbativity) and below (minimum value of $\mu$). 
These constraints translate into a range for $\tan\beta$:
\be
\frac 23 \left( \frac{1-a_{B_\mu}}{a_\mu a_{B_\mu}}\right)^2 \frac{\pi^4\, \Lambda_D^2\, \eta}{\alpha^2_t\alpha_3^2\, n_G\Lambda_G^2\, \log^2{M\over m_{\tilde t}}}
\lesssim\tan^2\beta\lesssim
\frac 54 \left( \frac{1-a_{B_\mu}}{a_{B_\mu}}\right)^2 \frac{\pi^2 \alpha_2^2}{\alpha_t\alpha_1 \alpha_3^2 \, \log{M\over m_{\tilde t}} \log{M\over m_{H_d}}}
\ee
where we have used the upper bound coming from the FI term.
For a fixed value of $\eta=m_Z^2/m^2_{\tilde t}=1\%$ we plot the bands for the allowed values of $\tan\beta$ in Fig.~\ref{figtanmax}, for the same parameter choices of Fig.~\ref{figmumax}. The upper bound, obtained by requiring positive square masses for the slepton doublets, is independent of $\Lambda_D/\Lambda_G$. The lower bound, on the other hand, scales as $\Lambda_D/\Lambda_G$ and both limits are approximately inversely proportional to $\log M$. We stress that since in lopsided gauge mediation $\tan\beta$ is simply controlled by the ratio of two Yukawas $\lambda_d$ and $\lambda_u$, its moderately large value is natural, as it does not require any additional tuning or cancellation among different contributions.\newline

Since the dependence of both $\mu$ and $\tan\beta$ on the UV parameters $x$, $y$, $\xi$, and $\theta$, turns out to be quite involved, we perform a scanning of such parameter space and calculate, for each point, the value of $\mu/M_1$ and $\tan\beta$. The results are shown in Fig.~\ref{scatmutan}. The anticorrelation between $\tan\beta$ and $\mu$ follows simply from the fact that their product depends on a combination of the variables which is bounded on the domain of the scatter plot. As observed in the right panel of Fig.~\ref{scatmutan}, relaxing the perturbativity bound on $\lambda_d$, that is imposing a cutoff lower than $M_{GUT}$, populates the region at larger values of $\tan\beta$ and $\mu/M_1$.

\section{Higgs mass and fine-tuning\label{tuning}}

Consistency with experimental data requires the Higgs boson mass to lie above the LEP bound $m_h>115$ GeV. This is a well-known source of tuning for supersymmetric theories in general and for those employing gauge mediation in particular.  As explained in the introduction, lopsided gauge mediation makes no exception to this. What it  does, however, is to hide behind this unavoidable tuning the one needed to accommodate the large $B_\mu$ typically found in models where all the soft terms are calculable. In sect.~3 we presented the possibly simplest implementation of lopsided gauge mediation and we observed its main phenomenological features: a large $m_{H_d}$ and light higgsinos and sleptons. 

LEP constrains the higgsino mass term to be bigger than about 100 GeV. Although this is never an issue in ordinary gauge mediation, it can become the most important source of tuning in the setup we are describing, as we already pointed out below eq.~(\ref{eta1}). As shown in the previous section, the ratio $\mu/M_1$ is fixed once and for all when the parameters appearing in eq.~(\ref{newmodel}) are chosen. $\Lambda_G$ is then fixing the overall normalization of the spectrum. 
To a very good approximation\footnote{A-term contributions to the stop masses, coming from the new couplings of the higgses to the messengers, are always irrelevant.} the mass of the lightest Higgs boson depends only on $\tan\beta$ and $m_{\tilde{t}}$, the latter being fixed by  $\Lambda_G$, $n_G$, $M$, as in ordinary gauge mediation. Once the ratio $\mu/M_1$ is fixed there is thus a maximal value of $\eta=m_Z^2/m^2_{\tilde t}$ such that the experimental bounds on $\mu$ and $m_h$ are simultaneously satisfied.

\begin{figure}[t]
 \begin{center}
     \includegraphics[width=10cm]{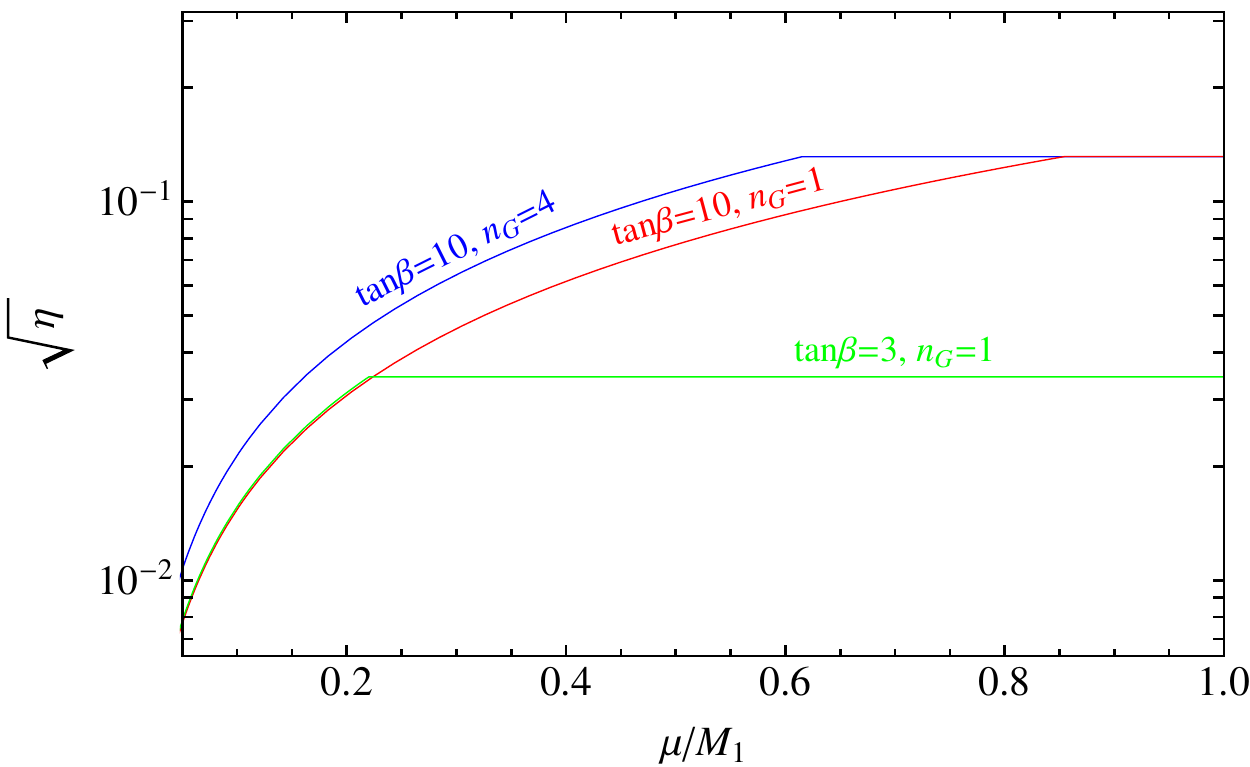}
 \end{center}
 \caption{Value of $\sqrt\eta$ as defined in the text for $M=10^9$\,GeV and various choices of $\tan\beta$, $n_G$.}
 \label{lambdastar}
\end{figure}

In Fig. \ref{lambdastar} we display the behaviour of this maximal $\sqrt\eta$ as a function of $\mu/M_1$ for definite values of $n_G$ and $\tan\beta$. 
When $\mu/M_1$ approaches 1, $\eta$ becomes independent of $\mu$: we are in the ordinary situation in which $\eta$ is fixed by the bound on the Higgs mass. Moving towards smaller values of $\mu/M_1$
we reach a point where the LEP bound on $\mu$ determines $\eta$, and the curve starts to fall as $(\mu/M_1)$.

The goal of lopsided gauge mediation is fully reached only if the lower bound on $\mu$ does not introduce additional fine-tuning with respect to the one required to lift the Higgs boson mass above the LEP experimental limit. This happens for models living on the flat (or nearly flat) part of the curves in Fig.~\ref{lambdastar}.

Figure~\ref{scatmutan} shows that, for a large fraction of the points, the ratio $\mu/M_{1}$ is not particularly large, especially for larger $\tan\beta$. The request that the tuning is dominated by the Higgs mass constraint severely restricts the allowed points. Another way to describe the situation can be extracted from eq.~(\ref{eta1}). Fixing $\eta$ to the value required by the Higgs mass is not always sufficient to guarantee that the lower bound on $\lambda_d$ is smaller than the upper bound from perturbativity. A further increase in tuning (obtained by decreasing $\eta$) is then required.

Figure~\ref{lambdastar} shows that, at small values of $\tan\beta$, lopsided gauge mediation requires no further tuning than ordinary gauge mediation. In this case, however, the overall degree of fine tuning is rather severe. For large values of $\tan\beta$, the overall tuning decreases, although lopsided gauge mediation can make the situation slightly worse. However, it should be noted that the effect is rather modest. For instance, the value of  $\eta$ at $\mu/M_1\approx 0.3$ and $\tan\beta =10$ does not significantly differ from the plateau at larger values of $\mu/M_1$. 
Moreover it should be remarked that, while  a moderate value of $\tan \beta$ generically requires an extra tuning of order $1/\tan\beta$ in the  case where all soft terms are comparable, in lopsided gauge mediation the structural hierarchies between $\mu$, the sfermion masses, and $m_A^2$ automatically imply a sizeable $\tan\beta$ without additional tuning.
This shows that lopsided gauge mediation can give spectra with tunings comparable to the ordinary case, although its phenomenological features are quite distinct.

\section{Other consequences of a heavy pseudoscalar}
\label{sect:otherheavy}

 The special spectrum we are dealing with -- a very heavy $H_d$ (several TeV) together with light higgsinos and left-handed sleptons (around 100 GeV) -- is expected to imprint a peculiar pattern of deviations in the running of gauge couplings with respect to a more typical supersymmetric spectrum. 
 The relevant equations needed to compute the new threshold effects are given in Appendix B.

\begin{figure}[t]
  \begin{center}
    \includegraphics[width=7.5cm]{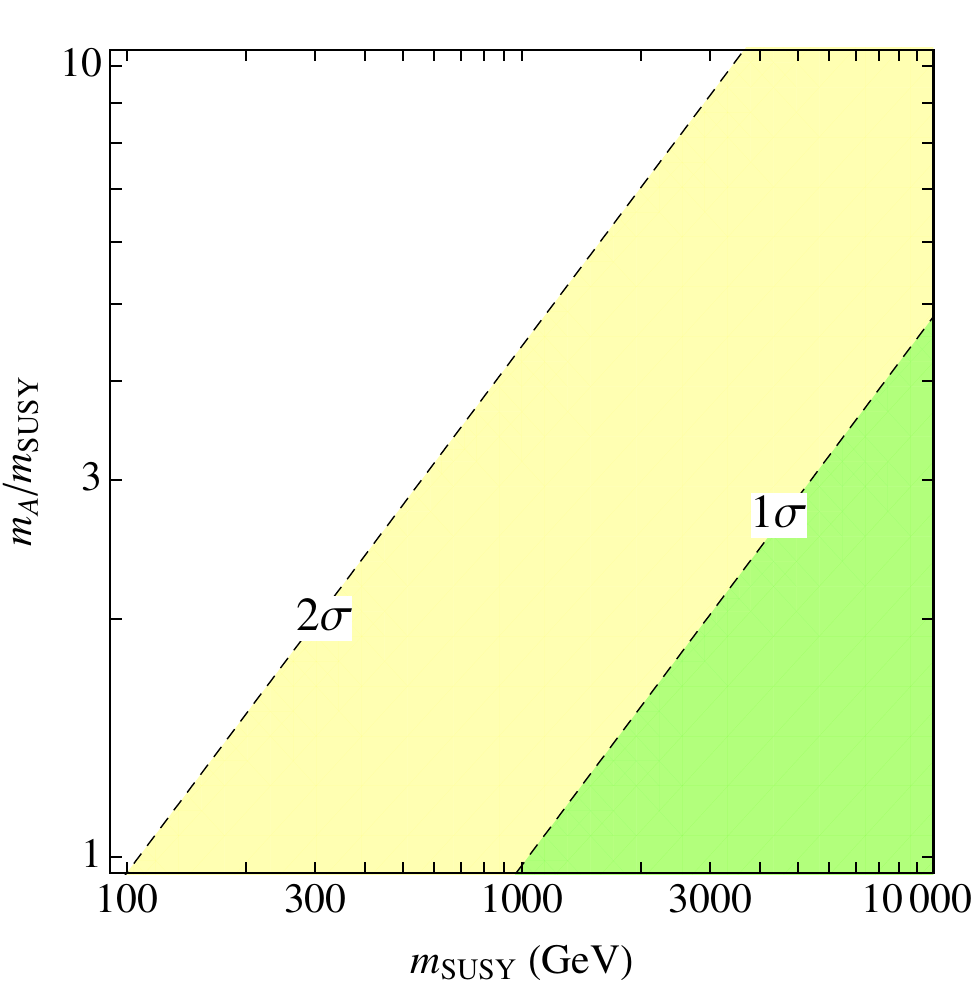}
  \end{center}
  \caption{Prediction of $\alpha_3(m_Z)$ including the $m_A$ threshold as explained in the text. In green (yellow) the $1\sigma$ ($2\sigma$) allowed range for $\delta\alpha_3/\alpha_3$, where $|\delta\alpha_3/\alpha_3|_{\rm exp}=0.006$.}
  \label{fig:alphas}
\end{figure}

Let us consider, for concreteness, only two thresholds between $m_Z$ and $M_{GUT}$, besides the messenger scale:
$m_A$, where the $H_d$ doublet is integrated out, and $m_{SUSY}<m_A$, where all other superpartners
except the bino, the higgsinos and the left-handed sleptons live. These latter particles are assumed to sit at $m_Z$.
Figure \ref{fig:alphas} shows the contours for the relative variation of the prediction for $\alpha_3(m_Z)$ with 
respect to the experimental measurement $\alpha_3(m_Z)=0.1184\times(1\pm 0.006)$~\cite{PDG}. Note that a heavy pseudoscalar gives a negative contribution to $\alpha_3(m_Z)$ which tends to spoil unification. On the other hand an improvement is obtained by splitting the higgsinos and the left-handed sleptons with respect to the rest of the supersymmetric spectrum.
\newline

\begin{figure}[t]
  \begin{center}
    \includegraphics[width=9cm]{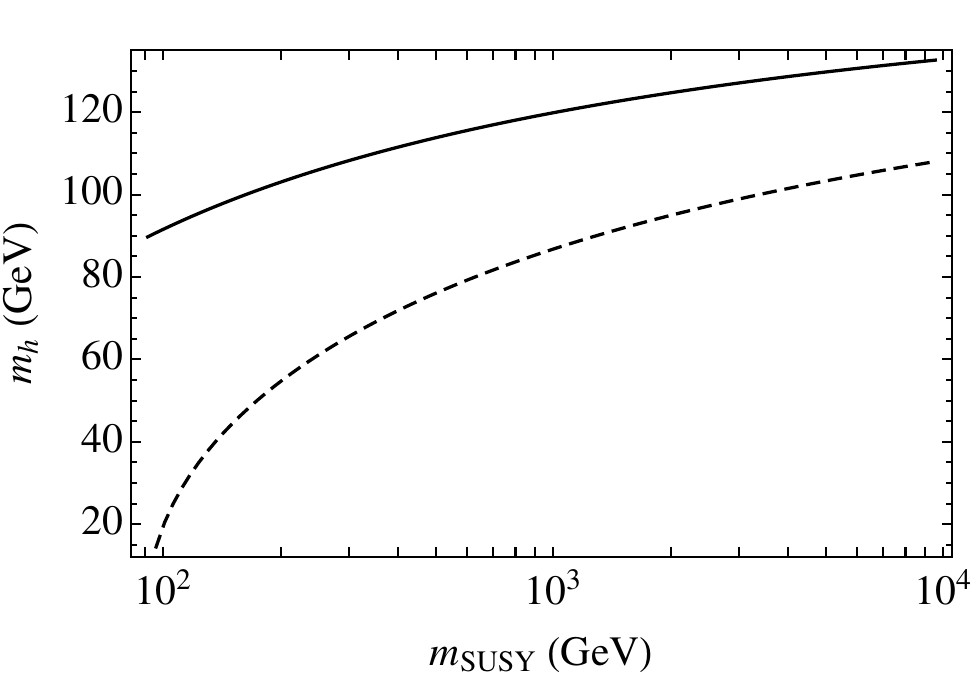}
  \end{center}
  \caption{\small Higgs boson mass at the weak scale for $\tan\beta=1$ (dashed), $\tan\beta=10$ (full), assuming all the superpartners to degenerate at the scale $m_{SUSY}$.
  }
  \label{CWhiggs}
\end{figure}

Another possible effect of a large $m_A$ is on the value of the light Higgs boson mass. The effect is analogous, but smaller, to the one obtained by integrating out the two stops and can be treated in the same way, using the renormalization group improved effective potential~\cite{effectivepotential}. Notice that it is consistent to single out this effect among others which are numerically of the same order, as it represent the leading contribution to $m_h$ proportional to $m_A$.
It is convenient to redefine the two doublets $H_{u,d}$ in terms of two fields 
\bea
\epsilon H_d^*=\cos\beta h -\sin\beta H,\\\nn
H_u=\sin\beta h+\cos\beta H,
\eea
where $\tan\beta=v_u/v_d$, so that only $h$ gets a vacuum expectation value. $H$ is identified with the heavy Higgs doublet and is integrated out at the scale $m_A$. Here supersymmetry fixes the boundary condition for the Higgs boson quartic 
\be
V(h)=-m^2|h|^2+\lambda|h|^4,\qquad \lambda (m_A)=\frac{g^2+g'^{2}}{8}\cos^22\beta.
\ee
We then assume all the other superpartners to be degenerate at the scale $m_{SUSY}<m_A$. Since $H$ is not present in the effective theory the quartic coupling at the  scale $m_{SUSY}$  is not fixed to its supersymmetric form but has an extra contribution
\be\label{extraquartic}
\lambda(m_{SUSY})=\lambda (m_A)+\delta\lambda\\
\ee
where $\delta\lambda$ is of the order $(g^2/4\pi)^2\log m_A/m_{SUSY}$. We can thus calculate the RG improved effective potential at the scale $M_Z$ using the new boundary condition in eq.~(\ref{extraquartic}) at the scale $m_{SUSY}$, running it down to the weak scale using the SM RGE equations. After this we obtain the Higgs boson mass as $m_h^2=4v^2\lambda$.
For all the values of $\tan\beta$ and $m_A$ which are relevant for our model this contribution to $m_h$ turns out to be always smaller than 1 GeV and thus negligible. In Fig.~\ref{CWhiggs} we show the value of the Higgs mass at the weak scale assuming all the superpartners to be degenerate at a scale $m_{SUSY}$.



\section{Collider Phenomenology}
In gauge mediation the lightest supersymmetric particle (LSP) is always the gravitino. However, in contrast with the ordinary case, in lopsided gauge mediation the two typical possibilities for the next-to-lightest supersymmetric particle (NLSP) are the higgsino, more or less mixed with the bino depending on the ratio $\mu/M_1$, or a sneutrino\footnote{The slepton doublet is splitted by the SU(2) $D$-terms of 10-20 GeV. $A_\tau$ terms are never relevant enough to invert the mass hierarchy between $\tilde\tau$ and $\tilde\nu_\tau$, so that the NLSP is always the sneutrino. The three sneutrino flavors can be considered degenerate for all practical purposes.} whose lightness is the result of the FI term induced by the large $m_{H_d}$ value. 

The identity of the NLSP depends on the parameters choice. In the majority of the parameter space we find a higgsino NLSP, as a result of the small $\mu$ of lopsided gauge mediation. Indeed to have a sneutrino lighter than the Higgsino one has to pick a choice of parameters $\lambda_d,n_G,\Lambda_D$ very close to the extreme values allowed by the experimental limits on the sleptons. We also notice that  typically this corresponds to a value of $\lambda_d$ that does not remain perturbative up the GUT scale, thus a sneutrino NLSP might be suggestive of some other new physics happening well below the GUT scale.

The generic spectrum of \lgm  have squarks heavier than 1.5 TeV and gluinos typically heavier than 2 TeV, which results in only a few fb of cross-section for the production of colored states at the LHC with 14 TeV of center of mass energy (LHC14) and negligible cross-section at the LHC with 7 TeV (LHC7). Thus, the production of electroweak sparticles in the cascade of the colored ones is not very abundant. On the other hand, the generic lightness of the higgsinos and the sleptons, can give sizable rates for the direct production  in  Drell-Yan processes mediated by electroweak bosons, independently of the mass of the colored states.

After its production any supersymmetric state decays promptly until the NLSP state is reached. The NLSP decay length is then determined by
\be
L\approx 10^{-2}\,{\rm cm}\left(\frac{100\,{\rm GeV}}{m_{NLSP}}\right)^5\left(\frac{\sqrt{F}}{100\,{\rm TeV}}\right)^4,
\ee
where $F$ is the scale appearing in the goldstino superfield $X_{NL}$ (which can be larger than the mass splitting among the messenger fields). The lower bound on the Higgs boson mass requires $\Lambda_G = k F/M> 10$\,TeV. The absence of tachyonic messengers, $\sqrt{k F}< M$ or equivalently $\Lambda_G< M$, thus implies a lower bound of roughly 10\,TeV on $\sqrt F$ if $k=\mathcal O(1)$\footnote{$k=1$ will be assumed in the following.}. In this range and with $m_{NLSP}\approx 100$\,GeV, values of $\sqrt{F}$ smaller than $100$\,TeV give rise to prompt NLSP decays, while when $\sqrt{F}$ is larger than a few times $10^3$\, TeV the NLSP decay takes place most of the time outside the detector. For intermediate values of the supersymmetry breaking scale the decay occurs via a displaced vertex.

\subsection{Higgsino NLSP spectrum}
The typical spectrum with a higgsino NLSP is characterized by a triplet of states at the bottom of the spectrum, the two neutral and the charged higgsinos, a neutral bino-like state parametrically heavier than the higgsinos and a yet heavier wino-like triplet.
The slepton doublets are lighter than the singlets and are typically lighter than the bino-like state. The typical spectrum in the less tuned region of the model looks as follows:

\be m_{\chi^0_1}\lesssim m_{\chi^\pm}\lesssim m_{\chi^0_2}<m_{\chi^0_3}\,,\quad m_{\tilde{\nu}} \lesssim m_{\tilde{l}_L}<m_{\chi^0_3}\,,\quad m_{\chi_4^0}\simeq 2 m_{\chi_3^0}\, \label{higgsinoNLSPspectrum}
\ee
\be
m_{\chi_1^0}\simeq \mu \lesssim 150 \gev\,,\quad m_{\chi^0_3}>400\gev\,,\quad m_{\tilde{q}}\gtrsim 1.5 \tev\tab m_{\tilde{g}}\gtrsim 2\tev.
\ee
Disregarding the signals from the production of electroweak states in the cascades of colored objects, the characteristic signals arise from the Drell-Yan production of charginos, neutralinos and sleptons and their decay to final states with many leptons:
\begin{itemize}
\item charged and neutral sleptons promptly decaying in final states with one lepton and one chargino or neutralino:
\be
 \tilde{\ell}^\pm \to \ell^\pm \chi_{1,2}^0\,,\,\chi^\pm\nu \tab \tilde{\nu}\to \chi_{1,2}^0\nu\,,\,\chi^\pm\ell^\mp \,;
\ee
\item light chargino and neutralino states, either directly produced or coming from the decay of the sleptons, decaying, through off-shell vector bosons or sleptons, in leptonic final states with a $\chi_1^0$:

\be
\chi_2^0\to \ell_i^\pm \ell_i^\mp \chi_1^0\tab    \chi_2^0\to \nu \bar{\nu} \chi_1^0\tab \chi^\pm \to \ell^\pm \nu \chi_1^0\,. \label{chi2decayll}
\ee
\end{itemize}
Altogether the resulting lepton-rich final states are: 
\bea
pp \textrm{ or } p\bar{p} &\to& \tilde{\ell}^\pm \tilde{\nu} \to 3\ell\, 2\chi_1^0\,\, \nu \textrm{ or } 2\nu \tab 5\ell\, 2\chi_1^0\nu\,, \label{slepPsnu}\\
pp \textrm{ or } p\bar{p} &\to& \tilde{\ell}^\pm \tilde{\ell}^\mp \to 6\ell\, 2\chi_1^0 \tab 4\ell\, 2\chi_1^0 \,2\nu\,,\\ 
pp \textrm{ or } p\bar{p} &\to& \tilde{\nu} \tilde{\nu} \to 4\ell\, 2\chi_1^0\, 2\nu \,,\label{sneuDY}\\
pp \textrm{ or } p\bar{p} &\to& \chi^\pm \chi_{2}^0  \to 3\ell\, 2\chi_1^0\,\, \nu\,, \label{chneu} \\
pp \textrm{ or } p\bar{p} &\to& \chi^\pm \chi^\mp \to 2\ell\, 2\chi_1^0 \,2\nu\,,\\ 
pp \textrm{ or } p\bar{p} &\to& \chi_2^0\chi_1^0 \to 2\ell\, 2\chi_1^0 \label{chi2chi1}\,.
\eea
We neglected the $\chi^0_2\chi^0_2$ channel which has a negligible production cross section for a higgsino-like $\chi^0_2$. Focusing on events with at least three leptons, at the LHC7 we expect at most $\mathcal O(10\,{\rm fb})$ of cross section, for sleptons and higgsino masses not excluded by LEP. Therefore the observability of such signals seems challenging in the 2011-2012 run of the LHC.

The decay products of $\chi_2^0$ can be used to determine the higgsino-like nature of both $\chi_2^0$ and $\chi_1^0$, as the invariant mass distribution of the di-lepton system from the decay of the $\chi_2^0$ is sensitive to the composition of the two neutralinos. This is due to CP invariance which requires different intrinsic parities for the $\chi_2^0\chi_1^0$ pair in case they are higgsino-like or gaugino-like \cite{phenochi}. This allows, in principle, for testing the prediction of \lgm of a light higgsino-like neutralino. However, we have to remark that, differently from the case studied in refs.~\cite{phenochi}, \lgm does not benefit from a copious source of moderately boosted $\chi_2^0$ from the decay chains of the colored sparticles, and therefore the leptons relevant for this analysis have $p_T\sim m_{\chi_2^0}-m_{\chi_1^0}\sim 10 \gev$ which makes them detectable but rather soft. A quantitative assessment of the potential of the LHC to measure the details of the invariant mass distribution of such soft leptons would be needed before concluding that the higgsino nature of the light neutralinos of \lgm can be tested at the LHC. 

A marked difference between \lgm and ordinary gauge mediation is the presence of doublet sleptons which are lighter than the singlet ones. The light sneutrino can in principle be discovered through the processes in eqs.~(\ref{slepPsnu}) or (\ref{sneuDY}).
Moreover, a crucial feature for distinguishing the left-handed sleptons from the right-handed ones, is the relevance of charged currents in the interactions. 
These leads to off-shell $W$s as in the following processes:
\bea
\chi_{3,4}^{0} &\to & \ell_i^{\mp}\tilde{\ell}_i^\pm \to \ell_i^{\mp} \chi^\pm \nu_l\to \ell_i^{\mp} W^{\pm*} \chi_1^0 \nu_\ell\,,\label{sleptoncharginosignal}\\
\tilde{\nu}& \to & \ell^\pm \chi_1^\mp\to \ell^\pm \chi_1^0 W^{\mp*} \,, \label{charginosignalsneutrino}
\eea
which are flavor universal sources of charged leptons. The processes eq.~(\ref{sleptoncharginosignal}) and (\ref{charginosignalsneutrino}) lead to both opposite-sign same-flavour (OSSF) and opposite-sign opposite-flavor (OSOF) final states. The invariant mass distribution of these OS di-leptons displays features which are characteristic of their production process. 
Backgrounds are expected not only from SM processes but also from other supersymmetric production mechanisms, which yield  featureless OS and SS di-lepton invariant mass distribution with similar shapes. Therefore, to isolate eq.~(\ref{sleptoncharginosignal}) and (\ref{charginosignalsneutrino}) one can subtract the OSSF and SSSF di-lepton invariant mass distribution (or equivalently the OSOF and SSOF), so that the contribution of the backgrounds is expected to cancel. This leaves a distribution whose features, \emph{e.g.} end-points, are connected to the mass differences in the chain. Similar reasoning can be applied to discover a sneutrino NLSP, as discussed in the following section.



\bigskip
All signatures discussed so far do not rely on the fate of $\chi^0_1$, the NLSP. In the following we shall examine the three different cases of a collider-stable NLSP, a meta-stable NLSP, and a promptly decaying NLSP. Along with the discussion of the signatures we shall present the current limits from searches at the TeVatron.
 

In the first case, typical of high-scale models of supersymmetry breaking, the $\chi_1^0$ is a massive invisible particle.
The signals in eqs.~(\ref{slepPsnu})--(\ref{chi2chi1}) result in this case in multi-leptons and missing transverse energy. 
The TeVatron experiments  D0 \cite{Abazov:2009zi} and CDF \cite{Forrest:2009gm,Aaltonen:2008pv} searched for new physics in the tri-lepton channel from the process in eq.~(\ref{chneu}).  The searches optimize the cuts for a typical mSUGRA spectrum and put a bound at around 100 fb for the source of tri-lepton events. This bound is comparable with the cross-section of  the process eq.~(\ref{chneu}) in \lgm for $\mu\simeq100 \gev$. We believe, however, that this kind of search is not effective in our case because the events selected in these analyses must have at least one lepton with $p_{T}\gtrsim15 \gev$. While this requirement is typically fulfilled by the mSUGRA signal, this is not the case for \lgm as the degeneracy of  $\chi_{2}^{0},\chi_{1}^{0},\chi^{\pm}$ limits the hardness of the leptons from  the process in eq.~(\ref{chneu}) \footnote{The search \cite{Aaltonen:2008my} performed a similar analysis on the final states $\mu^+\mu^-\mu^-$, $e^+\mu^-\mu^-$ and $\mu^+\mu^-e^-$ using a thresholds for $p_{T}$ as low as 5 GeV and observed 1 event in 0.96 ${\rm fb}^{-1}$ of data with 0.4 expected. Also in this case the limits are given only for a spectrum that comes from the mSUGRA model in which $\chi_{1}^{0}$ is several tens of GeV lighter than the $\chi_{2}^{0}$ and the $\chi_{1}^{\pm}$. As such their results cannot be translated into a limit for our case.}.

The process in eq.~(\ref{slepPsnu}) is in principle a source of tri-lepton events likely to pass the selection cuts employed in the searches at D0 \cite{Abazov:2009zi} and CDF \cite{Forrest:2009gm,Aaltonen:2008pv}. The production cross-section for such process is however significantly lower than the bound attainable even with the final TeVatron's integrated luminosity, around 10 ${\rm fb}^{-1}$.

\bigskip

A meta-stable neutralino can leave displaced vertex signatures, leading to striking signals of low-scale supersymmetry breaking. While the leptonic signals described above are still usable, the displaced vertex signature is an additional characteristic signal.
In the case of a higgsino NLSP, the relevant decay widths are
\bea
\Gamma(\chi\to\gamma\widetilde G)&=&\frac{1}{2}(\sin\beta+ \epsilon \cos\beta)^2 \cos^2\theta_W\sin^2\theta_W\left(\frac{m_Z}{M_1}\right)^2\left(\frac{m_\chi^5}{16\pi F^2}\right),\\
\Gamma(\chi\to Z\widetilde G)&=&\frac{1}{4}(\sin\beta+ \epsilon \cos\beta)^2 \left(1-\frac{m_Z^2}{m_\chi^2}\right)^4\left(\frac{m_\chi^5}{16\pi F^2}\right),\\
\Gamma(\chi\to h\widetilde G)&=&\frac{1}{4}(\sin\beta- \epsilon \cos\beta)^2 \left(1-\frac{m_h^2}{m_\chi^2}\right)^4\left(\frac{m_\chi^5}{16\pi F^2}\right),
\eea
where the $\epsilon$ parameter is the sign of the rephasing invariant combination $\mu M_{\tilde\lambda} B_\mu^*$.

The minimal gauge mediation scenario motivated a lot of effort on the signal $\gamma\gamma+\slashed{E}_T$, which is the dominant channel for a bino-like NLSP, \emph{i.e.} for  $M_1 \lesssim \mu$. In our case we expect the decays into $Z$ and Higgs bosons to dominate over the photon channel.
Displaced vertices from the decay $Z\to e^{-}e^{+}$ have been searched at the D0 experiment \cite{Abazov:2008zm}. The bounds strongly depend on the lifetime of the NLSP and the tightest bound from D0 is around 1 pb for $c\tau\simeq 0.2\, m$ \footnote{D0 searched, in addition, for displaced production of resonant $b\bar{b}$ pairs. Unfortunately the search \cite{Abazov:2009ik} considered only the displaced production of $b\bar{b}$ pairs with invariant mass of a few tens of GeV originating from the chain decay of a heavier resonance, \emph{e.g.} $s \to \eta \eta \to b\bar{b}b\bar{b}$ . Their result is not applicable to our case as the $b\bar{b}$ invariant mass of interest for the \lgm case is around $m_{Z}$ or $m_{h}$.}. In \lgm only electroweak particles are accessible at the TeVatron, yielding a total cross-section for $\chi_{1}^{0}$ production at most of about 1 pb for $\mu\gtrsim 100 \gev$.
Therefore only a very limited portion of the parameter space close to the LEP experimental bounds on $\mu$  and $c\tau\sim 0.2\, m$ is excluded by this search. 

Reference~\cite{meadelhc} considers the displaced decay of a neutralino taking place inside the ATLAS detector. The studied signatures are $\chi^0_1\rightarrow Z\widetilde G\rightarrow (e^+e^-)(\mu^+\mu^-)(jj)+\slashed{E}_T$. Neutralino decays to final states containing a Higgs boson can be assimilated to those with a hadronic $Z$ decay.  For LHC7, considering the  electroweak production of a meta-stable higgsino-like NLSP of mass 250 GeV,
ref.~\cite{meadelhc} claims that for decay lengths going from roughly $10^{-1}$ to $10^5$\,mm (corresponding to a range in $\sqrt F$ from a few hundreds to a few thousands TeV) at least a few signal events should be observable with 1 ${\rm fb}^{-1}$ of integrated luminosity. Thus, in the most natural portion of the parameter space of \lgm with $100 \tev \lesssim \sqrt{F}\lesssim 1000 \tev $, we expect that a few signal events will be produced in the 2011-2012 run of the LHC.

\bigskip

In the case of a promptly decaying $\chi^0_1$ there are further final state tracks originating from the primary vertex. 
The interesting channels for \lgm are the prompt decays $\chi_1^0\to Z/H+ \tilde G$, where the gravitino escapes the detector. Ref.~\cite{meadetevatron} used the results of non-dedicated  searches from TeVatron and estimated the limits on a pure higgsino NLSP. The photonic decay $\chi_1^0\to \gamma+ \tilde G$ does not put in our case any significant bound. On the contrary ref.~\cite{meadetevatron} finds that the CDF search \cite{highPTZ} for a high-$p_{T}$ $Z$ boson plus missing transverse energy can exclude  $\mu<150\gev$   for $M_{1,2}\gg \mu$ using the full integrated luminosity of TeVatron data. If this result is confirmed by the experimental collaborations a significant portion of the most natural part of the parameter space of \lgm with $\sqrt{F}\lesssim 100 \tev$ would be excluded.

\subsection{Sneutrino NLSP spectrum}
For particular choices of the parameters of \lgm the sneutrino can be lighter than the higgsino, resulting in a sneutrino NLSP spectrum. In these cases the spectrum will be very similar to the one in eq.~(\ref{higgsinoNLSPspectrum}) but with the inversion of the higgsino and the sneutrino at the bottom. Although the sneutrino is typically heavier than the higgsino at a generic point of the parameter space of \lgm, the sneutrino NLSP spectrum deserves special study due to its distinctive features. 

\bigskip

All supersymmetric particle production will eventually contribute to a pair of sneutrinos in the final state. These will decay, either promptly or not, in neutrinos and gravitinos,  acting always as sources of missing transverse energy.
Once produced, the higgsino states decay dominantly to sleptons via two-body decays
\be
\chi_{1,2}^{0}\to \ell^{\mp} \tilde{\ell}^{\pm}\,,\, \nu\tilde{\nu}\tab \chi_{1}^{\pm}\to \ell^{\pm} \tilde{\nu},\,\nu\tilde{\ell}^{\pm}\,,
\ee
and sub-dominantly via three-body decays
\be
\chi_{2}\to \chi_{1}^{0} Z^{*}\,,\, \chi^{\pm} W^{\mp,*} \tab \chi^{\pm} \to \chi_{1}^{0} W^{\pm,*} \,.
\ee
The charged sleptons, produced either directly or from the decay of a higgsino, decay into an off-shell $W$ and a sneutrino
\be
\tilde{\ell}\to \tilde{\nu}W^{*}\,.
\ee
Altogether the production and decay of higgsinos and sleptons lead to multi-leptons signal as: 
\bea
pp \textrm{ or } p\bar{p} &\to& \chi_{2}^{0}\chi_{1}^{0} \to 4 \ell+\slashed E_T\, \\
pp \textrm{ or } p\bar{p} &\to& \chi_{1,2}^{0} \chi_{1}^{\pm} \to 3\ell +\slashed E_T\label{trileptonsneutrino}\,,\\
pp \textrm{ or } p\bar{p} &\to& \chi_{1}^{+}\chi_{1}^{-} \to \ell^{-}\ell^{+}+\slashed E_T\,,\\
pp \textrm{ or } p\bar{p} &\to& \tilde{\ell}^{+}\tilde{\ell}^{-} \to \ell^{-}\ell^{+}+\slashed E_T\,.
 \eea
Comparing with the higgsino NLSP case we notice that these processes typically yield signals with fewer charged leptons with respect to those in eqs.~(\ref{slepPsnu})--(\ref{chi2chi1}).
Another difference is the possibility to pair produce the NLSPs with a non-vanishing cross sectio through the reaction
\be\label{snusnu}
pp \to \tilde{\nu}\tilde{\nu}\,.
\ee
This results in an invisible final state which can in principle be observed thanks to the emission of QCD initial state radiation. The resulting cross section (few fb) certainly requires large luminosity for an observation\footnote{The production of sneutrino at TeVatron is less than $\mathcal{O}(20)$~fb, which is significantly less than the sensitivity attainable with  the search \cite{Aaltonen:2008hh} using the final luminosity of about 10 ${\rm fb}^{-1}$.}.

\bigskip

The TeVatron searches \cite{Abazov:2009zi,Forrest:2009gm,Aaltonen:2008pv} are sensitive to the signal in eq.~(\ref{trileptonsneutrino}). Differently from the case of higgsino NLSP spectrum, the hardness of the leptons is controlled by the mass difference between the higgsino and the sleptons which, being in principle a free parameter, can be sufficiently large to yield hard leptons that pass the selection.
Therefore we have a bound of about 100 fb for the production cross-section of $\chi_{1,2}^{0}\chi_{1}^{\pm}$. This cross-section corresponds however to a higgsino whose mass lies very close to the experimental limit on the sleptons. This leads to soft leptons and thus invalidates the potential limit from refs.~\cite{Abazov:2009zi,Forrest:2009gm,Aaltonen:2008pv}.

\bigskip

The presence of a sneutrino NLSP leads to peculiar flavor and charge correlations in the multi-lepton final states. These originates from the $SU(2)$ charge of the NLSP and have been studied in detail in  refs.~\cite{katz1, katz2}.
The interesting phenomenon resides in the decay chain
\be\label{chainkatz}
\chi_{1,2}^{0}\rightarrow \ell^\pm\tilde\ell^{\mp}\rightarrow \ell^\pm W^{\mp*}\tilde \nu.
\ee
When the the virtual $W$ boson decays leptonically (\ref{chainkatz}) results in a final state with (at least) two leptons of opposite sign whose flavor is uncorrelated. This is in sharp contrast with the ordinary situation where the lightest slepton is an $SU(2)$ singlet and the chains of eq.~(\ref{chainkatz}) are absent. In that case opposite sign lepton pairs are always of the same flavor, as they arise from the decay of a heavier neutralino into the NLSP through an intermediate slepton. For a sneutrino NLSP spectrum with colored particles at the TeV, the authors of ref.~\cite{katz1, katz2} consider the relevant backgrounds to multi-lepton events and show how it is possible to reveal the presence of the process in eq.~(\ref{chainkatz}) looking at the differences between the same-sign and the opposite-sign di-lepton invariant mass distributions. Their subtraction reveal edges which are characteristic of the chain eq.~(\ref{chainkatz}).  Unfortunately the studies of ref.~\cite{katz1, katz2} are not immediately applicable to our spectrum, as the colored particles in our setup are well above the TeV. A dedicated study would thus be needed to conclude that the discovery of the flavor and charge correlations typical of a sneutrino NLSP are within the reach of the LHC.

\section{Conclusions}
\label{Conclusions}

In this paper we have expanded on the mechanism first proposed in ref.~\cite{csakietal} and considered a new class of models of gauge-mediated supersymmetry breaking (dubbed lopsided gauge mediation), where a single fine-tuning is sufficient both for evading the Higgs mass experimental limit
and for accomodating the large $B_\mu$ found in models where $\mu$ and $B_\mu$ originate at the same loop order.

We used an explicit model to analyze the distinctive features of our generic setup:
small $\mu$,  large $m_{H_d}$, light left-handed sleptons, and moderate to large $\tan\beta$. 
We find, in particular, that the perturbativity constraint on the higgs-messenger couplings puts rather
 stringent upper bounds on the ratio $\mu/M_1$. Depending on the value of $\tan\beta$, this 
 may or may not require an increase of the fine-tuning with respect to ordinary gauge mediation,
 where the value of $\mu$ is automatically large compared to the experimental limit, being fixed by 
 eq.~(\ref{casei}).

Although we have restricted ourselves to  gauge-mediated supersymmetry breaking,
the same mechanism can be  implemented in the case where the anomaly-mediated contribution to the soft masses is dominant.
In particular, one possibility is to extend the realization of ref.~\cite{gaugomaly} with the addition of two chiral singlets $S, \bar S$. The  superpotential couplings of the higgses to the messengers
can be chosen as in eq.~(\ref{newmodel}), while the supersymmetry-breaking mass terms for the singlets
get generated by K\"ahler potential couplings  to the conformal compensator $\varphi$ of the kind  
 $(\varphi^\dag/\varphi) (c_S S^2+ c_{\bar S}\bar S^2+2c_{S\bar S} S \bar S)$.
In this way, a viable Higgs sector with a large $H_d$ soft mass can be achieved.

The phenomenology of lopsided gauge mediation presents crucial differences with respect to ordinary gauge
mediation. It is characterized by light higgsinos (close to their experimental bounds),  large pseudo-scalar Higgs mass (in the several TeV range) and light left-handed sleptons. We have briefly discussed some possible signatures of this new class of models deserving experimental search.

\section*{Acknowledgements}
The work of A.D.S. and R.R. is supported by the Swiss National Science Foundation under 
contract 200021-125237.
The work of R.F. and D.P. is supported by the Swiss National Science Foundation under contract 200021-116372.

\appendix
\section{Effective K\"ahler}
\label{app:kahler}

In this appendix we provide some details of the calculation of the parameters of
the Higgs sector in eqs.~(\ref{mHfromKahler})--(\ref{At}).
The superpotential is described by eqs.~(\ref{extendedGMSB})--(\ref{Xparam}).
The messengers fields $D, \bar D, S, \bar S$ are integrated out at one loop giving rise to the effective
K\"ahler potential \cite{effectivekahler}
\be
K_{\rm eff}=-{1\over 32\pi^2}\Tr\left[
\mathcal{M}^\dag\mathcal{M}\log{\mathcal{M}^\dag\mathcal{M}\over \Lambda^2}
\right]\,,
\ee
where $\mathcal{M}$ is the field-dependent messenger mass matrix and $\Lambda$ is some UV
cutoff scale.
We then compute the eigenvalues of $\mathcal{M}^\dag \mathcal{M}$ and after some manipulations
we arrive at
\be
K_{\rm eff}=\mathcal{Z}_u H_u^\dag H_u+\mathcal{Z}_d H_d^\dag H_d+
(\mathcal{Z}_{ud} H_u H_d+\textrm{h.c.})\,,
\ee
where we kept only the quadratic terms in the propagating fields
and we have defined
\bea
\mathcal{Z}_{u,d}&=&
-{|\lambda_{u,d}|^2\over 16\pi^2}
\left[
{|X_S|^2\over |X_S|^2-|X_D|^2}\log{|X_S|^2\over \Lambda^2}
-{|X_D|^2\over |X_S|^2-|X_D|^2}\log{|X_D|^2\over \Lambda^2}
\right]\\
\mathcal{Z}_{ud}&=& -{\lambda_u\lambda_d \over 16\pi^2}{X_S^\dag X_D^\dag\over |X_S|^2-|X_D|^2}\log{\left|\frac{X_S}{X_D}\right|^2}\,.
\eea
To fix the sign conventions we will use the lagrangian
\be
\mathcal L = -m_{H_u}^2|H_u|^2-m_{H_d}^2|H_d|^2-\sum_iA_i H_i\partial_{H_i}W-B_\mu H_uH_d+\int d\theta^2\,\mu H_u H_d.
\ee
The terms proportional to $|H_{u,d}|^2$ in the K\"ahler potential generate  contributions to the Higgs masses and the A-terms
\bea
m_{H_{u,d}}^2&=&-\left.{\partial\over \partial\theta^2}{\partial\over \partial\bar\theta^2}
\log{\mathcal Z}_{u,d}\right\vert_{\theta^2=\bar\theta^2=0}\\
A_{u,d}&=&\left.{\partial\over \partial\theta^2}
\log{\mathcal Z}_{u,d}\right\vert_{\theta^2=\bar\theta^2=0}\,,
\eea 
while the mixed term $H_u H_d$ generates $\mu$ and $B_\mu$
\bea
\mu&=&\left.{\partial\over \partial\bar\theta^2}
{\mathcal Z}_{ud}\right\vert_{\theta^2=\bar\theta^2=0}\\
B_\mu&=&-\left.{\partial\over \partial\theta^2}{\partial\over \partial\bar\theta^2}
{\mathcal Z}_{ud}\right\vert_{\theta^2=\bar\theta^2=0}\,.
\eea 
These expressions are easily evaluated and lead to the results given in the text in eqs.~(\ref{mHfromKahler})--(\ref{loopfunctions2}).
\newline

With the previous technique it is also straightforward to derive the effective K\"ahler potential in the model with vanishing $B_\mu$. With the same notation we used above
\bea
\mathcal{Z}_{u,d}&=&
-{|\lambda_{u,d}|^2\over 16\pi^2}
|X|^2\left[
\frac{2 M_D^4-3 M_D^2M_S^2+M_S^4-M_D^2M_S^2\log M_D^2/M_S^2}{2M_D^2(M_D^2-M_S^2)^2}
\right],\\
\mathcal{Z}_{ud}&=& -{\lambda_u\lambda_d \over 16\pi^2} X^\dagger \left[\frac{M_S(M_S^2-M_D^2+M_S^2\log M_D^2/M_S^2)}{(M_D^2 -M_S^2)^2}\right] \,.
\eea
The structure of the potential imply the vanishing of both $B_\mu$ and the $A-$terms at one-loop.

The fact that $m^2_{H_u}$ and $m^2_{H_d}$ are generically non zero in a model where $B_\mu$ vanishes due to an $R-$symmetry can be simply understood. In order for the $R$-symmetry to be exact the charge of $X$ has to be fixed to 2. This means that the K\"ahler potential can only be a function of the invariant combination $X X^\dagger$. Furthermore since $X$ has no scalar VEV, no mass can vanish in the limit $X\to 0$ and the K\"ahler potential is an analytic function around that point. The most general form of $\mathcal Z_{u,d}$ is thus
\be
\mathcal Z_{u,d}=f(M,M^\dagger)+g(M,M^\dagger) X X^\dagger,
\ee
and, unless $g$ vanishes accidentaly, the soft masses are generated.

\section{Tree-level threshold corrections to unification}
We supply here the necessary formulas to calculate the corrections to the low energy value of the strong coupling constant due to tree-level thresholds.

The running gauge couplings at low scale can  be written in terms of the unified coupling $\alpha_G$, the grand scale $M_{GUT}$ and all the various thresholds
\be
\frac{1}{\alpha_i}=\frac{1}{\alpha_G}+\sum_{i=1}^{n-1}\frac{b^{(i)}}{4\pi}\log\frac{M_{i}^2}{M_{i+1}^2}
\ee
where $M_1=M_{GUT}$ and $M_n=M_Z$. The formula can be written as
\be\label{running}
\frac{1}{\alpha_i}=\frac{1}{\alpha_G'}+\frac{b^{(i)}_{MSSM}}{4\pi}\log\frac{M_{GUT}}{\mu^2}+X_i(\mu)
\ee
$\mu$ is an arbitrary scale and the $X_i$ contains all thresholds that do not depend on $M_{GUT}$.
Finally we absorbed in $\alpha_G'$ a possible unified contribution to the running, for instance coming from the presence,  at some scale $M$, of messenger fields in a unified representation of the gauge group. If this is the case the relation between $\alpha_G$ and $\alpha_G'$ is
\be
\frac{1}{\alpha_G'}=\frac{1}{\alpha_G}+\frac{n_G}{4\pi}\log\frac{M_{GUT}^2}{M^2};
\ee
where $n_G$ is the Dynkin index of the representation of the messengers.

With these definitions one obtains a prediction for $\alpha_3$ as a function of $\alpha_1$ and $\alpha_2$ at low energy
\bea
\log\frac{M_{GUT}^2}{\mu^2}&=&\frac{4\pi}{b_{12}}\left\{ \frac{1}{\alpha_1(M_Z)}-\frac{1}{\alpha_2(M_Z)}-\left[ X_1(\mu)-X_2(\mu)\right] \right\} \\
\frac{1}{\alpha_3(M_Z)}-\frac{1}{\alpha_3^{(0)}(M_Z)}&=&-X_1(\mu)\frac{b_{23}}{b_{21}}-X_2(\mu)\frac{b_{13}}{b_{12}}+X_3(\mu)
\eea
where
\be
b_{ij}=b^{(i)}_{MSSM}-b^{(j)}_{MSSM}
\ee
\be
\frac{1}{\alpha_3^{(0)}}=\frac{1}{\alpha_1(M_Z)}\frac{b_{23}}{b_{21}}+\frac{1}{\alpha_2(M_Z)}\frac{b_{13}}{b_{12}}.
\ee

For the specific setup described in the text we give the $b$ functions in the three relevant regions: $\mu>m_A$ (region $I$) where the whole MSSM spectrum is propagating, $m_A>\mu>m_{SUSY}$ (region $II$) where a combination of the two Higgs doublets has been integrated out, $m_{SUSY}>\mu>m_Z$ (region $III$) where only the SM states together with higgsinos and left-handed sleptons are left in the spectrum,
\bea
b^{I}&=&(33/5, 1, -3)\\
b^{II}&=&(13/2,5/6, -3) \\
b^{III}&=&(24/5,-2,-7).
\eea

\end{document}